\documentclass[showpacs,twocolumn,superscriptaddress]{revtex4} % twocolumn
\usepackage{epsfig,amsmath,amssymb,amsbsy,graphicx}
\usepackage[cp1251]{inputenc}
\usepackage[english]{babel}  %

\begin{document}

\title{Bose-Einstein condensation in a mixture of interacting Bose and Fermi particles} %
%\author{Yu.M.\,Poluektov}
\author{Yu.M. Poluektov}
\email{yuripoluektov@kipt.kharkov.ua} %
\affiliation{National Science Center ``Kharkov Institute of Physics
and Technology'', Akhiezer Institute for Theoretical Physics, 61108
Kharkov, Ukraine} %
\affiliation{V.N.\,Karazin Kharkov National University, 61077
Kharkov, Ukraine}
\author{A.A. Soroka} %
\affiliation{National Science Center ``Kharkov Institute of Physics
and Technology'', Akhiezer Institute for Theoretical Physics, 61108
Kharkov, Ukraine}

\begin{abstract}
A self-consistent field model for a mixture of Bose and Fermi
particles is formulated. There is explored in detail the case of a
delta-like interaction, for which the thermodynamic functions are
obtained, and Bose-Einstein condensation of interacting particles in
the presence of the admixture of fermions is studied. It is shown
that the admixture of Fermi particles leads to reducing of the
temperature of Bose-Einstein condensation and smoothing of features
of thermodynamic quantities at the transition temperature. As in the
case of a pure Bose system, in the state of a mixture with
condensate the dependence of the thermodynamic potential on the
interaction constant between Bose particles has a nonanalytic
character, so that it proves impossible to develop the perturbation
theory in the magnitude of interaction of Bose particles.
\newline%
\end{abstract}
\maketitle %  \cite{}

\section{Introduction}
The phenomenon of Bose-Einstein condensation \cite{Bose,Einstein}
was used by F.\,London \cite{London} and  Tisza \cite{Tisza} to
explain the phenomenon of superfluidity of liquid helium discovered
by Kapitsa \cite{Kapitsa} and Allen \cite{Allen}. Yet the model of
an ideal gas is too simple to explain properties of dense systems in
which the interparticle interaction plays a substantial role, the
fact that was pointed out by Landau \cite{Landau1}. But, as was
demonstrated in experiments on neutron scattering in the superfluid % $^4$He
$^4$He \cite{Woods,BKKP}, Bose-Einstein condensate exists also in
the presence of the interparticle interaction. A new splash of
interest to the phenomenon of superfluidity and its relationship to
condensation is associated with the discovery about 20 years ago of
Bose-Einstein condensation in atomic gases of alkali metals confined
in magnetic \cite{PitStr,Pethick} and laser traps \cite{Pit1}.

In the middle of last century there appeared a possibility of
obtaining the isotope $^3$He in significant quantities, which
enabled to start a systematic study of the properties of the quantum
systems consisting of bosons and fermions --\, %
$^3$He\,--\,$^4$He solutions \cite{Eselson}. Landau and Pomeranchuk
considered \cite{Landau3} on a qualitative level the behavior of the
admixture of $^3$He in the superfluid $^4$He at a low concentration
of fermions and showed that the admixture particles do not
participate in the superfluid motion. In work by Bardeen, Baym, and
Pines \cite{BBP} it was shown how the form of the effective pair
potential of interaction between the admixture quasiparticles can be
reconstructed based on the comparison to experimental data. A review
of theoretical approaches for description of the properties of
$^3$He\,--\,$^4$He solutions and a discussion of the problem of
superfluidity therein are given in works \cite{Bashkin,Kagan}. %

Presently, along with a research on condensation of Bose atoms in
magnetic and laser traps, there is a growing interest in studying in
such traps of the properties of gases which atoms obey the Fermi
statistics \cite{Turlapov}, and of their mixtures. Theoretically a
mixture of bosons and fermions at zero temperature in external field
was considered in \cite{Molmer}. A study of the properties of
mixtures of Fermi and Bose particles from different perspectives was
performed in relatively recent papers \cite{Watanabe,Adhikari,Maruyama,Noda}. %
Note, however, that the analysis of influence of the Fermi admixture
on the character in itself of the phase transition into the state
with Bose-Einstein condensate has actually received little
attention.

In work \cite{P1} it was proposed a relatively simple model of
Bose-Einstein condensation of interacting particles, in which the
transition into the state of lowest energy with a macroscopic number
of particles is accounted for in a similar way as is done within the
Bose-Einstein model of an ideal gas \cite{Einstein}. There is a
number of difficulties in the theory of Bose-Einstein condensation
of an ideal gas, in particular connected with the fact that the
description of condensation requires to fix the chemical potential
which is considered as an independent thermodynamic variable above
the transition temperature. Besides that, below the condensation
temperature the pressure proves to be a function of only temperature
and does not depend on the density. As a consequence the isobaric
heat capacity \cite{P2} and the isothermal compressibility become
infinite, in contradiction with general principles of thermodynamics
according to which the heat capacities have to tend to zero as
temperature approaches zero \cite{LL}. A direct indication of the
limitation of the model of Bose-Einstein condensation in an ideal
gas is that the fluctuation of the number of particles in the phase
with condensate becomes infinite \cite{LL}. Accounting for the
interaction between particles enables to eliminate these
difficulties \cite{P1} and ensure the fulfillment of the correct
thermodynamic relations in the state with condensate as well. \newpage %

In this paper, on the basis of the model of condensation of
interacting particles, which was proposed in \cite{P1}, we consider
the influence of Fermi particles on Bose-Einstein condensation, with
the number of admixture particles not being assumed small. %
At first, the self-consistent field equations for a mixture of
interacting bosons and fermions in the absence of Bose-Einstein
condensate and for arbitrary pair interaction potentials are
formulated in general form. The case of a delta-like interaction
between particles is explored in detail and formulas for the
thermodynamic potential, equation of state, entropy, heat
capacities, isothermal and adiabatic compressibilities are obtained.
The transition into the state with condensate is described in a
similar way as it was done for a pure Bose system in \cite{P1}, at
that the chemical potentials of both components of a mixture remain
``good'' thermodynamic variables and the calculated thermodynamic
quantities in the phase with condensate satisfy all general
requirements. The conditions of stability of the spatially uniform
states are considered. The results of some specific calculations of
the temperature dependencies of pressure, heat capacities and
compressibilities at different concentrations of Fermi particles are
presented. It is shown that adding the Fermi admixture leads to
reducing of the condensation temperature and smoothing of features
of thermodynamic quantities at this temperature. It should be noted
that this model is applicable for description of mixtures of
sufficiently dilute gases, and what concerns the properties of
$^3$He\,--\,$^4$He solutions one can hope only for their qualitative
description.

\section{Self-consistent field equations for a normal fermion-boson system} %
Let us consider a system of volume $V$, which is a mixture of a
large number $N_F$ of fermions with spin $s=1/2$ and mass $m_F$ and
a large number of spinless bosons $N_B$ with mass $m_B$. In the
second-quantized representation the system is described by the field
operators: the fermion $\Psi(q)$ where $q\equiv({\bf r},\sigma)$,
$\sigma=1,2$ is the fermion spin index, and the boson one $\Phi({\bf r})$. %
The field operators satisfy respectively the standard
anticommutation and commutation relations \cite{LL2}. %
The Hamiltonian of the fermion-boson system has the form
\begin{equation} \label{01}
\begin{array}{l}
\displaystyle{%
  H=H_F+H_B+H_{F\!B}, %
}%
\end{array}
\end{equation}
where
\begin{equation} \label{02}
\begin{array}{l}
\displaystyle{%
  H_F=\int\!\! dqdq'\Psi^+(q)H_F(q,q')\Psi(q')\,+ %
} \vspace{2mm}\\ %
\displaystyle{%
  \hspace{7mm} +\frac{1}{2}\int\!\! dqdq' \Psi^+(q)\Psi^+(q')U_F({\bf r},{\bf r}')\Psi(q')\Psi(q), %
}%
\end{array}
\end{equation}
\begin{equation} \label{03}
\begin{array}{l}
\displaystyle{%
  H_B=\int\!\! d{\bf r}d{\bf r}'\Phi^+({\bf r})H_B({\bf r},{\bf r}')\Phi({\bf r}')\,+ %
} \vspace{2mm}\\ %
\displaystyle{%
  \hspace{7mm} +\frac{1}{2}\int\!\! d{\bf r}d{\bf r}' \Phi^+({\bf r})\Phi^+({\bf r}')U_B({\bf r},{\bf r}')\Phi({\bf r}')\Phi({\bf r}) %
}%
\end{array}
\end{equation}
-- the Hamiltonians of Fermi and Bose particles with taking into
account their pair interaction,
\begin{equation} \label{04}
\begin{array}{l}
\displaystyle{%
  H_{F\!B}=\int\!\! dqd{\bf r}'\Psi^+(q)\Phi^+({\bf r}')U_{F\!B}({\bf r},{\bf r}')\Phi({\bf r}')\Psi(q) %
}%
\end{array}
\end{equation}
-- the Hamiltonian of the pair interaction between bosons and
fermions. Here
\begin{equation} \label{05}
\begin{array}{l}
\displaystyle{%
  H_F(q,q')=-\frac{\hbar^2}{2m_F}\Delta\,\delta(q-q')-\mu_F\delta(q-q')\,+ %
} \vspace{1mm}\\ %
\displaystyle{%
  \hspace{48mm} +\,U_F({\bf r})\delta(q-q'), %
}\vspace{2mm}\\ %
\displaystyle{%
  H_B({\bf r},{\bf r}')=-\frac{\hbar^2}{2m_B}\Delta\,\delta({\bf r}-{\bf r}')-\mu_B\delta({\bf r}-{\bf r}')\,+ %
}\vspace{1mm}\\ %
\displaystyle{%
  \hspace{48mm} +\,U_B({\bf r})\delta({\bf r}-{\bf r}'), %
}
\end{array}
\end{equation}
$\mu_F, \mu_B$ -- the chemical potentials of the Fermi and Bose systems, %
$U_F({\bf r},{\bf r}')$, $U_B({\bf r},{\bf r}')$, $U_{F\!B}({\bf r},{\bf r}')$ %
-- the energies of interaction of Fermi, Bose particles and between
Fermi and Bose particles, $U_F({\bf r})$, $U_B({\bf r})$ -- the
energies of interaction with external fields. The operators of the
number of fermions and bosons are given by the formulas:
\begin{equation} \label{06}
\begin{array}{l}
\displaystyle{%
  N_{F}=\sum_\sigma\!\int\!\Psi^+(q)\Psi(q) d{\bf r}, \quad  %
  N_{B}=\int\!\Phi^+({\bf r})\Phi({\bf r}) d{\bf r}.  %
}%
\end{array}
\end{equation}

Let us derive the self-consistent field equations for a mixture in
the absence of Bose-Einstein condensate in a formulation that
was developed earlier for Fermi and Bose systems in works \cite{P3,P4,P5}. %
For this purpose, the many-particle Hamiltonian (\ref{01}) is
divided into the self-consistent $H_0$ and the correlation $H_C$
parts, so that
\begin{equation} \label{07}
\begin{array}{l}
\displaystyle{%
  H=H_0+H_C.  %
}%
\end{array}
\end{equation}
The self-consistent Hamiltonian is chosen quadratic in the field
operators in the form
\begin{equation} \label{08}
\begin{array}{l}
\displaystyle{%
  H_0=\int\!\!dqdq'\Psi^+(q)\big[H_F(q,q')+W_F(q,q')\big]\Psi(q')\,+ %
} \vspace{2mm}\\ %
\displaystyle{%
  \hspace{6mm} +\int\!\!d{\bf r}d{\bf r}'\Phi^+({\bf r})\big[H_B({\bf r},{\bf r}')+W_B({\bf r},{\bf r}')\big]\Phi({\bf r}') + E_0. %
}
\end{array}
\end{equation}
Here there enter the self-consistent fields $W_F(q,q')$, $W_B({\bf
r},{\bf r}')$ and the non-operator energy $E_0$, which have to be
determined later. The correlation Hamiltonian contains all terms
which do not enter into the self-consistent Hamiltonian:
\begin{equation} \label{09}
\begin{array}{l}
\displaystyle{%
  H_C=\frac{1}{2}\int\!\! dqdq' \Psi^+(q)\Psi^+(q')U_F({\bf r},{\bf r}')\Psi(q')\Psi(q)\,+ %
} \vspace{2mm}\\ %
\displaystyle{%
  \hspace{7mm} +\frac{1}{2}\int\!\! d{\bf r}d{\bf r}' \Phi^+({\bf r})\Phi^+({\bf r}')U_B({\bf r},{\bf r}')\Phi({\bf r}')\Phi({\bf r})\,+ %
}\vspace{2mm}\\ %
\displaystyle{%
  \hspace{7mm} +\int\!\! dqd{\bf r}'\Psi^+(q)\Phi^+({\bf r}')U_{F\!B}({\bf r},{\bf r}')\Phi({\bf r}')\Psi(q)\,- %
}\vspace{2mm}\\ %
\displaystyle{%
  \hspace{7mm} -\int\!\! dqdq'\Psi^+(q)W_F(q,q')\Psi(q')\,- %
}\vspace{2mm}\\ %
\displaystyle{%
  \hspace{7mm} -\int\!\! d{\bf r}d{\bf r}'\Phi^+({\bf r})W_B({\bf r},{\bf r}')\Phi({\bf r}')-E_0. %
}
\end{array}
\end{equation}
It should be stressed that the decomposition of the full Hamiltonian
into the self-consistent (\ref{08}) and the correlation (\ref{09})
parts does not change its form. The self-consistent Hamiltonian
(\ref{08}) can be written in the form similar to Hamiltonian of free
particles. To this end, let us introduce the quasiparticle operators
of fermions $\gamma_j^+,\gamma_j$ and bosons $\eta_i^+,\eta_i$ which
are connected with the field operators by the relations
\begin{equation} \label{10}
\begin{array}{ll}
\displaystyle{%
  \Psi(q)=\sum_j \phi_j(q)\,\gamma_j, \quad \Psi^+(q)=\sum_j \phi_j^*(q)\,\gamma_j^+, %
} %
\end{array}
\end{equation}\vspace{-4mm}
\begin{equation} \label{11}
\begin{array}{ll}
\displaystyle{%
  \Phi({\bf r})=\sum_i \varphi_i({\bf r})\,\eta_i, \quad \Phi^+({\bf r})=\sum_i \varphi_i^*({\bf r})\,\eta_i^+. %
} %
\end{array}
\end{equation}
The one-particle functions in (\ref{10}), (\ref{11}) are supposed to
obey the conditions of orthonormality and completeness:
\begin{equation} \label{12}
\begin{array}{ll}
\displaystyle{%
  \int\!\!dq\phi_j^*(q)\phi_{j'}(q)=\delta_{jj'},\quad \sum_j\phi_j^*(q)\phi_j(q')=\delta(q-q'), %
} \vspace{2mm}\\ %
\displaystyle{%
  \int\!\!d{\bf r}\varphi_i^*({\bf r})\varphi_{i'}({\bf r})=\delta_{ii'},\quad \sum_i\varphi_i^*({\bf r})\varphi_i({\bf r}')=\delta({\bf r}-{\bf r}'), %
}
\end{array}
\end{equation}
so that the transformations (\ref{10}) and (\ref{11}) are canonical
and the quasiparticle operators $\gamma_j^+,\gamma_j$ and $\eta_i^+,\eta_i$ %
satisfy the anticommutation relations for fermions and the
commutation relations for bosons \cite{LL2}. The one-particle
functions in (\ref{10}) and (\ref{11}) are subject to the equations of self-consistency %
\begin{equation} \label{13}
\begin{array}{l}
\displaystyle{%
  \int\!\!dq'\big[H_F(q,q')+W_F(q,q')\big]\phi_j(q')=\varepsilon_{Fj}\phi_j(q), %
}
\end{array}
\end{equation}\vspace{-4mm}
\begin{equation} \label{14}
\begin{array}{l}
\displaystyle{%
  \int\!\!d{\bf r}'\big[H_B({\bf r},{\bf r}')+W_B({\bf r},{\bf r}')\big]\varphi_i({\bf r}')=\varepsilon_{Bi}\varphi_i({\bf r}). %
}
\end{array}
\end{equation}
It follows from (\ref{13}),(\ref{14}) that in terms of the
quasiparticle operators the self-consistent Hamiltonian has the form
\begin{equation} \label{15}
\begin{array}{l}
\displaystyle{%
  H_0=\sum_j\varepsilon_{Fj}\gamma_j^+\gamma_j + \sum_i\varepsilon_{Bi}\eta_i^+\eta_i\,+ E_0. %
}
\end{array}
\end{equation}
The quantities $\varepsilon_{Fj},\, \varepsilon_{Bi}$ have the
meaning of the energies of Fermi and Bose quasiparticles. %
Thus, the Hamiltonian (\ref{15}) is a sum of the Fermi
$H_{0F}=\sum_j\varepsilon_{Fj}\gamma_j^+\gamma_j$ and the Bose %
$H_{0B}=\sum_i\varepsilon_{Bi}\eta_i^+\eta_i$ Hamiltonians, %
so that $H_0=H_{0F}+H_{0B}+E_0$. Note that the Fermi and Bose
Hamiltonians commutate.

Let us define the statistical operator
\begin{equation} \label{16}
\begin{array}{ll}
\displaystyle{%
  \hat{\rho}_0=\exp\beta(\Omega-H_0),
} %
\end{array}
\end{equation}
where $\beta=1/T$, the constant
$\Omega=-T\ln\!\big[\textrm{Sp}\exp(-\beta H_0)\big]$ is determined
from the normality condition $\textrm{Sp}\hat{\rho}_0=1$ %
and has the meaning of the thermodynamic potential of the system in
the self-consistent field model:
\begin{equation} \label{17}
\begin{array}{ll}
\displaystyle{%
  \Omega=E_0-T\ln\textrm{Sp}\Big[e^{-\beta H_{0F}}e^{-\beta H_{0B}}\Big] = %
} \vspace{2mm}\\ %
\displaystyle{\hspace{3mm}%
  =E_0-T\sum_j\ln\left(1+e^{-\beta\varepsilon_{Fj}}\right) + T\sum_i\ln\left(1-e^{-\beta\varepsilon_{Bi}}\right).   %
}
\end{array}
\end{equation}
The average of an arbitrary operator $A$ in the self-consistent
field approximation is defined by the relation
\begin{equation} \label{18}
\begin{array}{ll}
\displaystyle{%
  \langle A \rangle \equiv \textrm{Sp}(\hat{\rho}_0 A).
} %
\end{array}
\end{equation}

The energy $E_0$ is found from the requirement of maximal proximity
of the approximating self-consistent Hamiltonian to the exact
Hamiltonian. Qualitatively this condition can be formulated as the
requirement of the quantity $I=\big|\big\langle H-H_0\big\rangle\big|$ %
being minimal. Hence it follows the usual condition of the mean
field theory
\begin{equation} \label{19}
\begin{array}{ll}
\displaystyle{%
  \big\langle H\big\rangle = \big\langle H_0\big\rangle. %
} %
\end{array}
\end{equation}
Therefore
\begin{equation} \label{20}
\begin{array}{l}
\displaystyle{%
  E_0=\frac{1}{2}\int\!\! dqdq'U_F({\bf r},{\bf r}')\big\langle\Psi^+(q)\Psi^+(q')\Psi(q')\Psi(q)\big\rangle\,+ %
} \vspace{2mm}\\ %
\displaystyle{%
  \hspace{7mm} +\frac{1}{2}\int\!\! d{\bf r}d{\bf r}' U_B({\bf r},{\bf r}')\big\langle \Phi^+({\bf r})\Phi^+({\bf r}')\Phi({\bf r}')\Phi({\bf r})\big\rangle\,+ %
}\vspace{2mm}\\ %
\displaystyle{%
  \hspace{7mm} +\int\!\! dqd{\bf r}'U_{F\!B}({\bf r},{\bf r}')\big\langle\Psi^+(q)\Phi^+({\bf r}')\Phi({\bf r}')\Psi(q)\big\rangle\,- %
}\vspace{2mm}\\ %
\displaystyle{%
  \hspace{7mm} -\int\!\! dqdq'W_F(q,q')\big\langle\Psi^+(q)\Psi(q')\big\rangle\,- %
}\vspace{2mm}\\ %
\displaystyle{%
  \hspace{7mm} -\int\!\! d{\bf r}d{\bf r}'W_B({\bf r},{\bf r}')\big\langle\Phi^+({\bf r})\Phi({\bf r}')\big\rangle. %
}
\end{array}
\end{equation}
Let us define the one-particle density matrices by the relations: %
\begin{equation} \label{21}
\begin{array}{ll}
\displaystyle{%
  \rho_F(q,q')=\big\langle\Psi^+\!(q')\Psi(q)\big\rangle = \sum_j\phi_j(q)\phi_j^*\!(q')f_{Fj}, %
} \vspace{2mm}\\ %
\displaystyle{%
  \rho_B({\bf r},{\bf r}')=\big\langle\Phi^+\!({\bf r}')\Phi({\bf r})\big\rangle = \sum_i\varphi_i({\bf r})\varphi_i^*\!({\bf r}')f_{Bi}.    %
}
\end{array}
\end{equation}
It is taken into account here that
$\langle\gamma_j^+\gamma_{j'}\rangle=f_{Fj}\delta_{j'j}$ and %
$\langle\eta_i^+\eta_{i'}\rangle=f_{Bi}\delta_{i'i}$, and the Fermi
and Bose distribution functions have the usual form
\begin{equation} \label{22}
\begin{array}{ll}
\displaystyle{%
  f_{Fj}=f(\varepsilon_{Fj})=\big(\!\exp\beta\varepsilon_{Fj}+1\big)^{-1},  %
}\vspace{3.5mm}\\ %
\displaystyle{%
  f_{Bi}=f(\varepsilon_{Bi})=\big(\!\exp\beta\varepsilon_{Bi}-1\big)^{-1}.    %
}
\end{array}
\end{equation}
Thus
\begin{equation} \label{23}
\begin{array}{l}
\displaystyle{%
  E_0=\frac{1}{2}\int\!\! dqdq'U_F({\bf r},{\bf r}')\times %
}\vspace{0mm}\\ %
\displaystyle{%
  \hspace{20mm} \times\big[\rho_F(q,q)\rho_F(q',q')-\rho_F(q',q)\rho_F(q,q')\big]\,+ %
}\vspace{2mm}\\ %
\displaystyle{%
  \hspace{7mm} +\frac{1}{2}\int\!\! d{\bf r}d{\bf r}' U_B({\bf r},{\bf r}')\times %
}\vspace{0mm}\\ %
\displaystyle{%
  \hspace{20mm} \times\big[ \rho_B({\bf r},{\bf r})\rho_B({\bf r}',{\bf r}')+\rho_B({\bf r}',{\bf r})\rho_B({\bf r},{\bf r}')\big]\,+ %
}\vspace{2mm}\\ %
\displaystyle{%
  \hspace{7mm} +\int\!\! dqd{\bf r}'U_{F\!B}({\bf r},{\bf r}')\rho_F(q,q)\rho_B({\bf r}',{\bf r}')\,- %
}\vspace{2mm}\\ %
\displaystyle{%
  \hspace{7mm} -\int\!\! dqdq'W_F(q,q')\rho_F(q',q)\,- %
}\vspace{2mm}\\ %
\displaystyle{%
  \hspace{7mm} -\int\!\! d{\bf r}d{\bf r}'W_B({\bf r},{\bf r}')\rho_B({\bf r}',{\bf r}). %
}
\end{array}
\end{equation}

The variation of the thermodynamic potential %
$\Omega=-T\ln\!\Big[\!\sum_n\!\big\langle n\big|e^{-\beta H_0}\big|n\big\rangle \Big]$ %
is equal to the averaged variation of Hamiltonian (\ref{08}): %
\begin{equation} \label{24}
\begin{array}{ll}
\displaystyle{%
  \delta\Omega=\frac{\sum_n\!\big\langle n\big|e^{-\beta H_0}\delta H_0\big|n\big\rangle}
               {\sum_n\!\big\langle n\big|e^{-\beta H_0}\big|n\big\rangle} =  % \vphantom{\Big)}
  \big\langle\delta H_0\big\rangle.
}
\end{array}
\end{equation}
From the requirement that the variations of the thermodynamic
potential with respect to the density matrices (\ref{21}) vanish
$\delta\Omega\big/\delta\rho_F(q,q')=0$, $\delta\Omega\big/\delta\rho_B({\bf r},{\bf r}')=0$, %
we obtain the form of the self-consistent potentials:
\begin{equation} \label{25}
\begin{array}{ll}
\displaystyle{%
  W_F(q,q')=-U_F({\bf r},{\bf r}')\rho_F(q,q')\,+ %
}\vspace{2mm}\\ %
\displaystyle{%
  \hspace{5mm} +\,\delta(q-q')\Bigg[\int\!\!dq''U_F({\bf r},{\bf r}'')\rho_F(q'',q'')\,+ %
}\vspace{0mm}\\ %
\displaystyle{%
  \hspace{28mm} +\int\!\!d{\bf r}''U_{F\!B}({\bf r},{\bf r}'')\rho_B({\bf r}'',{\bf r}'')\Bigg], %
}
\end{array}
\end{equation}\vspace{-3mm}
\begin{equation} \label{26}
\begin{array}{ll}
\displaystyle{%
  W_B({\bf r},{\bf r}')=U_B({\bf r},{\bf r}')\rho_B({\bf r},{\bf r}')\,+ %
}\vspace{2mm}\\ %
\displaystyle{%
  \hspace{5mm} +\,\delta({\bf r}-{\bf r}')\Bigg[\int\!\!d{\bf r}''U_B({\bf r},{\bf r}'')\rho_B({\bf r}'',{\bf r}'')\,+ %
}\vspace{0mm}\\ %
\displaystyle{%
  \hspace{28mm} +\int\!\!dq''U_{F\!B}({\bf r},{\bf r}'')\rho_F(q'',q'')\Bigg]. %
}
\end{array}
\end{equation}
Substitution of these expressions into (\ref{23}) gives
\begin{equation} \label{27}
\begin{array}{l}
\displaystyle{%
  E_0=-\frac{1}{2}\int\!\! dqdq'U_F({\bf r},{\bf r}')\times %
}\vspace{0mm}\\ %
\displaystyle{%
  \hspace{20mm} \times\big[\rho_F(q,q)\rho_F(q',q')-\rho_F(q',q)\rho_F(q,q')\big]\,- %
}\vspace{2mm}\\ %
\displaystyle{%
  \hspace{9mm} -\frac{1}{2}\int\!\! d{\bf r}d{\bf r}' U_B({\bf r},{\bf r}')\times %
}\vspace{0mm}\\ %
\displaystyle{%
  \hspace{20mm} \times\big[ \rho_B({\bf r},{\bf r})\rho_B({\bf r}',{\bf r}')+\rho_B({\bf r}',{\bf r})\rho_B({\bf r},{\bf r}')\big]\,- %
}\vspace{2mm}\\ %
\displaystyle{%
  \hspace{9mm} -\int\!\! dqd{\bf r}'U_{F\!B}({\bf r},{\bf r}')\rho_F(q,q)\rho_B({\bf r}',{\bf r}'). %
}
\end{array}
\end{equation}
The thermodynamic potential is a function of the temperature $T$ and
chemical potentials $\mu_F$, $\mu_B$, as well as a functional of the
one-particle density matrices $\rho_F(q,q')=\rho_F(q,q';T,\mu_F,\mu_B)$, %
$\rho_B({\bf r},{\bf r}')=\rho_B({\bf r},{\bf r}';T,\mu_F,\mu_B)$
which also depend on these quantities. But since the self-consistent
potentials (\ref{25}),(\ref{26}) were determined from the conditions
$\delta\Omega\big/\delta\rho_F(q,q')=0$ and $\delta\Omega\big/\delta\rho_B({\bf r},{\bf r}')=0$, %
then when taking the derivatives of the thermodynamic potential with
respect to temperature and chemical potentials one should take into
account only the explicit dependence of $\Omega$ on $T$, $\mu_F$, $\mu_B$. %
This ensures the fulfilment of all thermodynamic relations within
the self-consistent field model. The equations (\ref{13}),\,(\ref{14}) % \,--\,
together with the formulas for the self-consistent fields
(\ref{25}),\,(\ref{26}) describe the properties of a mixture of Bose
and Fermi particles in the normal state within the self-consistent field model. %
This model can be chosen as the main approximation, and based on it
there can be developed the quantum field perturbation theory where
the correlation Hamiltonian (\ref{09}) serves a perturbation \cite{P4,P5}. %
In this work we confine ourselves to consideration of a mixture only in the %
self-consistent field model.\vspace{-1mm}

\section{Spatially uniform state with point interaction}\vspace{-1mm}
Let us consider the spatially uniform nonmagnetic system in the
absence of external fields. The interaction potentials will be
assumed to have a delta-like form
\begin{equation} \label{28}
\begin{array}{ccc}
\displaystyle{%
  U_F({\bf r}-{\bf r}')=g_F\delta({\bf r}-{\bf r}'), \quad %
  U_B({\bf r}-{\bf r}')=g_B\delta({\bf r}-{\bf r}'),
}\vspace{4mm}\\ %
\displaystyle{%
  \hspace{0mm} U_{F\!B}({\bf r}-{\bf r}')=g_{F\!B}\delta({\bf r}-{\bf r}'). %
}
\end{array}
\end{equation}
In this case the dispersion laws of quasiparticles remain the same
as in an ideal gas
\begin{equation} \label{29}
\begin{array}{l}
\displaystyle{%
  \varepsilon_{Fk}=\frac{\hbar^2k^2}{2m_F} - \mu_{F*},\quad  %
  \varepsilon_{Bk}=\frac{\hbar^2k^2}{2m_B} - \mu_{B*},  %
}%
\end{array}
\end{equation}
but the chemical potentials are replaced by the effective chemical
potentials
\begin{equation} \label{30}
\begin{array}{l}
\displaystyle{%
  \mu_{F*}=\mu_F-\frac{1}{2}g_Fn_F-g_{F\!B}n_B,   %
}\vspace{4mm}\\ %
\displaystyle{%
  \mu_{B*}=\mu_B-2g_Bn_B-g_{F\!B}n_F.   %
}
\end{array}
\end{equation}
The appearance of the coefficients 1/2 before $g_F$ and 2 before
$g_B$ in (\ref{30}) is conditioned by taking into account the
exchange interaction. The dispersion laws of quasiparticles in an
interacting system change, if the nonlocal character of the
interaction between particles is being taken into account. In this
case the masses of free particles are replaced by their effective
masses \cite{PS}.

Similar to the case of the ideal quantum gases, all thermodynamic
functions can be expressed through the standard tabulated functions
\begin{equation} \label{31}
\begin{array}{l}
\displaystyle{%
  \Phi_s^{(\pm)}(t)=\frac{1}{\Gamma(s)}\int_0^{\!\infty} \frac{z^{s-1}\,dz}{e^{z-t}\pm 1}, %
}%
\end{array}
\end{equation}
where the plus sign refers to the Fermi particles and the minus sign
to the Bose particles, $\Gamma(s)$ is the gamma function. The
functions $\Phi_s^{(-)}(t)$ are defined for $t\leq 0$,
and for $t>0$ the integrals $\Phi_s^{(-)}(t)$ diverge. %
We will need functions (\ref{31}) with a half-integer index $s\geq 1/2$. %
At $t=0$ functions (\ref{31}) are expressed through the Riemann zeta function $\zeta(s)$: %
$\Phi_s^{(+)}(0)=(1-2^{1-s})\zeta(s)$ and $\Phi_s^{(-)}(0)=\zeta(s)$ if $s>1$, %
and at $t\rightarrow -0$\,\, $\Phi_{1/2}^{(-)}(t)\approx\sqrt{-\pi/t}$. %
Note that in the case $s>1$ there is a useful relation
$d\Phi_s^{(\pm)}(t)\big/dt=\Phi_{s-1}^{(\pm)}(t)$. The functions
$\Phi_s^{(\pm)}(t)$ monotonically increase with increasing $t$, and
at $t\rightarrow -\infty$ there holds for them the asymptotics
$\Phi_s^{(\pm)}(t)\approx e^t$. %

The numbers of fermions and bosons, as functions of chemical
potentials, are given by the formulas:
\begin{equation} \label{32}
\begin{array}{ll}
\displaystyle{%
  N_F=\frac{2V}{\Lambda_F^3}\Phi_{3/2}^{(+)}(t_F),\quad %
  N_B=\frac{V}{\Lambda_B^3}\Phi_{3/2}^{(-)}(t_B), %
}
\end{array}
\end{equation}
where the parameters $t_F=\beta\mu_{F*}$, $t_B=\beta\mu_{B*}$ are defined, and %
\begin{equation} \label{33}
\begin{array}{ll}
\displaystyle{%
  \Lambda_F\equiv\big(2\pi\hbar^2\big/m_FT\big)^{1/2},\quad %
  \Lambda_B\equiv\big(2\pi\hbar^2\big/m_BT\big)^{1/2} %
}
\end{array}
\end{equation}
are the de Broglie thermal wavelengths. Pay attention that, in
contrast to the case of ideal gases, the effective chemical
potentials themselves (\ref{30}) are connected with the number
densities of fermions and bosons, so that (\ref{32}) is a coupled
system of nonlinear equations.

%%%
The thermodynamic potential is given by expression % the
\begin{equation} \label{34}
\begin{array}{ll}
\displaystyle{%
  \Omega=E_0 - \frac{2TV}{\Lambda_F^3}\Phi_{5/2}^{(+)}(t_F)- \frac{TV}{\Lambda_B^3}\Phi_{5/2}^{(-)}(t_B). %
}
\end{array}
\end{equation}
Here $E_0$ is obtained from the formula (\ref{27}):
\begin{equation} \label{35}
\begin{array}{l}
\displaystyle{%
  \frac{E_0}{V}=-g_F\frac{n_F^2}{4}-g_Bn_B^2-g_{F\!B}n_Fn_B.   %
}
\end{array}
\end{equation}
Therefore
\begin{equation} \label{36}
\begin{array}{ll}
\displaystyle{\hspace{0mm}%
  \Omega=-V\Bigg[g_F\frac{n_F^2}{4}+g_Bn_B^2+g_{F\!B}n_Fn_B\, + %
}\vspace{1mm}\\ %
\displaystyle{%
  \hspace{25mm} + \frac{2T}{\Lambda_F^3}\Phi_{5/2}^{(+)}(t_F)+ \frac{T}{\Lambda_B^3}\Phi_{5/2}^{(-)}(t_B)\Bigg].   %
}
\end{array}
\end{equation}
The thermodynamic potential should be considered as a function of
the chemical potentials $\mu_F$, $\mu_B$ and temperature $T$, since
according to (\ref{32})
\begin{equation} \label{37}
\begin{array}{ll}
\displaystyle{%
  n_F=\frac{2}{\Lambda_F^3}\Phi_{3/2}^{(+)}(t_F),\quad %
  n_B=\frac{1}{\Lambda_B^3}\Phi_{3/2}^{(-)}(t_B). %
}
\end{array}
\end{equation}
The system of equations (\ref{37}) determines the densities as
functions of the chemical potentials and temperature:
$n_F=n_F(T,\mu_F,\mu_B)$ and $n_B=n_B(T,\mu_F,\mu_B)$.

Since the pressure $p=-\Omega\big/V$, it follows from (\ref{36}): %
\begin{equation} \label{38}
\begin{array}{ll}
\displaystyle{\hspace{0mm}%
  p=g_F\frac{n_F^2}{4}+g_Bn_B^2+g_{F\!B}n_Fn_B\, + %
}\vspace{1mm}\\ %
\displaystyle{%
  \hspace{20mm} + \frac{2T}{\Lambda_F^3}\Phi_{5/2}^{(+)}(t_F)+ \frac{T}{\Lambda_B^3}\Phi_{5/2}^{(-)}(t_B).   %
}
\end{array}
\end{equation}
The total energy is given by the formula
\begin{equation} \label{39}
\begin{array}{ll}
\displaystyle{\hspace{0mm}%
  E=E_0+2\sum_k(\varepsilon_{Fk}+\mu_F)f_{Fk} + \sum_k(\varepsilon_{Bk}+\mu_B)f_{Bk}, %
}
\end{array}
\end{equation}
so that
\begin{equation} \label{40}
\begin{array}{ll}
\displaystyle{\hspace{0mm}%
  \frac{E}{V}=g_F\frac{n_F^2}{4}+g_Bn_B^2+g_{F\!B}n_Fn_B\, + %
}\vspace{1mm}\\ %
\displaystyle{%
  \hspace{20mm} +\,3\frac{T}{\Lambda_F^3}\Phi_{5/2}^{(+)}(t_F)+ \frac{3}{2}\frac{T}{\Lambda_B^3}\Phi_{5/2}^{(-)}(t_B).   %
}
\end{array}
\end{equation}

The entropy can be obtained both by the combinatorial way
\begin{equation} \label{41}
\begin{array}{ll}
\displaystyle{\hspace{-2mm}%
 S = -2\sum_k\!\big[f_{Fk}\ln f_{Fk} +(1-f_{Fk})\ln(1-f_{Fk})\big]\,-
}\vspace{1mm}\\ %
\displaystyle{%
  \hspace{7mm} -\sum_k\!\big[f_{Bk}\ln f_{Bk} -(1+f_{Bk})\ln(1+f_{Bk})\big],   %
}
\end{array}
\end{equation}
and by means of the thermodynamic relation $S=-(\partial\Omega/\partial T)_{\mu_F,\mu_B}$: %
\begin{equation} \label{42}
\begin{array}{ll}
\displaystyle{\hspace{0mm}%
  S=\frac{2V}{\Lambda_F^3}\Bigg[\frac{5}{2}\Phi_{5/2}^{(+)}(t_F)-t_F\Phi_{3/2}^{(+)}(t_F)\Bigg] +  %
}\vspace{2mm}\\ %
\displaystyle{%
  \hspace{22mm} + \frac{V}{\Lambda_B^3}\Bigg[\frac{5}{2}\Phi_{5/2}^{(-)}(t_B)-t_B\Phi_{3/2}^{(-)}(t_B)\Bigg].   %
}
\end{array}
\end{equation}

Sometimes, instead of the number densities of particles of each
sort, it is more convenient to use the total density and
concentration. Let there is a total number of particles $N=N_F+N_B$
in the volume $V$, then the total density $n=N/V=n_F+n_B$. %
Let us define the concentrations of the Fermi and Bose particles:
\begin{equation} \label{43}
\begin{array}{ll}
\displaystyle{\hspace{-2mm}%
  x_F\equiv \frac{N_F}{N}=\frac{n_F}{n},\quad  %
  x_B\equiv \frac{N_B}{N}=\frac{n_B}{n}.
}
\end{array}
\end{equation}
In the following we will consider the Fermi component as an
admixture, setting $x_F\equiv x$ and $x_B\equiv 1-x$. Then
\begin{equation} \label{44}
\begin{array}{cc}
\displaystyle{\hspace{0mm}%
  N_F=Nx, \quad N_B=N(1-x), %
}\vspace{2mm}\\ %
\displaystyle{%
  n_F=nx, \quad n_B=n(1-x).   %
}
\end{array}
\end{equation}

One can use also the mass concentrations:
\begin{equation} \label{45}
\begin{array}{cc}
\displaystyle{\hspace{0mm}%
  M=M_F+M_B=m_FN_F+m_BN_B, %
}\vspace{2mm}\\ %
\displaystyle{%
  c_F\equiv c\equiv\frac{M_F}{M}=\frac{m_F x}{m_F x+m_B(1-x)},   %
}\vspace{2mm}\\ %
\displaystyle{%
  c_B\equiv 1-c\equiv\frac{M_B}{M}=\frac{m_B(1-x)}{m_F x+m_B(1-x)}.   %
}
\end{array}
\end{equation}
The mass densities are given by the formulas: %
$\rho\equiv M/V=n\big[m_Fc+m_B(1-c)\big]$, %
$\rho_F\equiv M_F/V=c\rho$, $\rho_B\equiv M_B/V=(1-c)\rho$. %

\section{Thermodynamics of a mixture in the absence of Bose-Einstein condensate}
In a system of noninteracting Bose particles at some temperature
$T_C$ the condensation in the momentum space takes place
\cite{Einstein}. In \cite{P1} it was proposed a relatively simple
generalization of the model with taking into account the interaction
between particles in a pure Bose system. The phenomenon of
Bose-Einstein condensation will also take place in a mixture of
fermions and bosons. At first, in this section, we consider
thermodynamics of a mixture of Fermi and Bose particles interacting
according to (\ref{28}), in the spatially uniform case at $T>T_C$,
when the condensate is absent. At high temperatures the effective
chemical potentials (\ref{30}) are negative and they increase with
decreasing temperature. At the temperature when the effective
chemical potential of Fermi particles becomes zero the system does
not have any peculiarities, and under the condition $\mu_{B*}=0$ the
condensation occurs in a mixture in the subsystem of Bose particles.
The temperature of Bose-Einstein condensation $T_C$ like in a Bose
gas is determined by the density of Bose particles:
\begin{equation} \label{46}
\begin{array}{ll}
\displaystyle{%
   n_B=\frac{\Phi_{3/2}^{(-)}(0)}{\big[\Lambda_B(T_C)\big]^3} = n(1-x). %
}
\end{array}
\end{equation}
At the fixed total density the condensation temperature decreases
with increasing the concentration of the admixture of Fermi
particles. Note that in $^3$He\,\,--\,$^4$He solutions the
temperature of the superfluid transition also decreases with
increasing the concentration of the admixture of Fermi particles
\cite{Eselson}. The thermodynamic potential, densities, pressure,
energy and entropy in the case under consideration are determined by
the formulas (\ref{36})\,--\,(\ref{38}), (\ref{40}) and (\ref{42}).
The parameters entering into these formulas in the phase without
condensate vary in the intervals $-\infty<t_B\le 0$ and $-\infty<t_F
< t_{F0}$, where $t_{F0}$ is determined by the equation
\begin{equation} \label{47}
\begin{array}{ll}
\displaystyle{%
   \frac{n_F}{2n_B}\left(\frac{m_B}{m_F}\right)^{\!3/2}=\frac{\Phi_{3/2}^{(+)}(t_{F0})}{\Phi_{3/2}^{(-)}(0)}. %
}
\end{array}
\end{equation}

Let us derive the heat capacities of a solution. To this aim we have
to calculate the differentials of the entropy, numbers of Fermi and
Bose particles, and also pressure:
\begin{equation} \label{48}
\begin{array}{cc}
\displaystyle{%
  dS=S\frac{dV}{V}+\frac{3}{2}S\frac{dT}{T}+\frac{2V}{\Lambda_F^3}\!\left( \frac{3}{2}\,\Phi_{3/2}^{(+)} - t_F\Phi_{1/2}^{(+)} \right)\!dt_F\,+ %
}\vspace{2mm}\\ %
\displaystyle{%
\hspace{34mm}  +\,\frac{V}{\Lambda_B^3}\!\left( \frac{3}{2}\,\Phi_{3/2}^{(-)} - t_B\Phi_{1/2}^{(-)}\right)\!dt_B,   %
}
\end{array}
\end{equation}\vspace{-0mm}
\begin{equation} \label{49}
\begin{array}{cc}
\displaystyle{%
  dN_F=N_F\frac{dV}{V}+\frac{3}{2}N_F\frac{dT}{T}+\frac{2V}{\Lambda_F^3}\Phi_{1/2}^{(+)}\,dt_F, %
}\vspace{3mm}\\ %
\displaystyle{%
dN_B=N_B\frac{dV}{V}+\frac{3}{2}N_B\frac{dT}{T}+\frac{V}{\Lambda_B^3}\Phi_{1/2}^{(-)}\,dt_B, %
}
\end{array}
\end{equation}\vspace{-0mm}
\begin{equation} \label{50}
\begin{array}{lllll}
\hspace{-0mm}
\displaystyle{%
  dp=\!\left[5\frac{\Phi_{5/2}^{(+)}}{\Lambda_F^3}+\frac{5}{2}\frac{\Phi_{5/2}^{(-)}}{\Lambda_B^3}\right]\!dT\,+  %
}\vspace{2mm}\\ %
\displaystyle{%
\hspace{5mm}  +\frac{2T}{\Lambda_F^3}\Phi_{3/2}^{(+)}dt_F +\frac{T}{\Lambda_B^3}\Phi_{3/2}^{(-)}dt_B\,+ %
}\vspace{2mm}\\ %
\displaystyle{%
\hspace{5mm}  +\left(g_F\frac{n_F}{2}+g_{F\!B}n_B\right)n_F\frac{dN_F}{N_F}\,+ %
}\vspace{2mm}\\ %
\displaystyle{%
\hspace{5mm}  +\big(2g_Bn_B +g_{F\!B}n_F\big)\,n_B\frac{dN_B}{N_B}\,- %
}\vspace{2mm}\\ %
\displaystyle{%
\hspace{5mm}  -\,2\left(g_F\frac{n_F^2}{4}+g_Bn_B^2 +g_{F\!B}n_Fn_B\right)\!\frac{dV}{V}. %
}
\end{array}
\end{equation}
Then at arbitrary conditions the heat capacity can be written in the form: %
\begin{equation} \label{51}
\begin{array}{lll}
\displaystyle{%
  C=T\frac{dS}{dT}= %
}\vspace{2mm}\\ %
\displaystyle{%
\hspace{2mm} = \!\frac{3}{2}S+\frac{ST}{V}\frac{dV}{dT}+\frac{2VT}{\Lambda_F^3}\!\left( \frac{3}{2}\,\Phi_{3/2}^{(+)} - t_F\Phi_{1/2}^{(+)} \right)\!\frac{dt_F}{dT}\,+ %
}\vspace{2mm}\\ %
\displaystyle{%
\hspace{30mm} + \frac{VT}{\Lambda_B^3}\!\left( \frac{3}{2}\,\Phi_{3/2}^{(-)} - t_B\Phi_{1/2}^{(-)}\right)\!\frac{dt_B}{dT}.   %
}
\end{array}
\end{equation}
As was remarked above, we consider the system with a fixed number of
particles of each sort in the volume $V$. In other words, the
concentration $x$ is assumed to be fixed. From the assumed condition
of constancy of the numbers of particles ($dN_F=dN_B=0$) we get the
relations:
\begin{equation} \label{52}
\begin{array}{cc}
\displaystyle{%
  \frac{1}{V}\frac{dV}{dT}=-\frac{3}{2}\beta - \!\frac{\Phi_{1/2}^{(+)}}{\Phi_{3/2}^{(+)}}\frac{dt_F}{dt},\,\,\, %
  \frac{1}{V}\frac{dV}{dT}=-\frac{3}{2}\beta - \!\frac{\Phi_{1/2}^{(-)}}{\Phi_{3/2}^{(-)}}\frac{dt_B}{dt}. %
}
\end{array}
\end{equation}
From here it follows:
\begin{equation} \label{53}
\begin{array}{ll}
\displaystyle{%
  \frac{\Phi_{1/2}^{(+)}}{\Phi_{3/2}^{(+)}}\frac{dt_F}{dt}=\frac{\Phi_{1/2}^{(-)}}{\Phi_{3/2}^{(-)}}\frac{dt_B}{dt}\equiv -\eta,\,\,\, %
  \frac{1}{V}\frac{dV}{dT}=-\frac{3}{2}\beta +\eta. %
}
\end{array}
\end{equation}
Using (\ref{53}), we obtain formulas at $x=const$ for the heat
capacity at a constant volume:
\begin{equation} \label{54}
\begin{array}{ll}
\displaystyle{\hspace{-2mm}%
  \frac{C_{V,x}}{N}=\frac{15}{4}\Bigg\{ x\Bigg[\frac{\Phi_{5/2}^{(+)}(t_F)}{\Phi_{3/2}^{(+)}(t_F)} - \frac{3}{5}\frac{\Phi_{3/2}^{(+)}(t_F)}{\Phi_{1/2}^{(+)}(t_F)} \Bigg]\,+ %
}\vspace{2mm}\\ %
\displaystyle{%
\hspace{15mm} +\, (1-x)\Bigg[\frac{\Phi_{5/2}^{(-)}(t_B)}{\Phi_{3/2}^{(-)}(t_B)} - \frac{3}{5}\frac{\Phi_{3/2}^{(-)}(t_B)}{\Phi_{1/2}^{(-)}(t_B)} \Bigg]\!\Bigg\}   %
}
\end{array}
\end{equation}
and the heat capacity at a constant pressure: % and the isobaric heat capacity:
\begin{equation} \label{55}
\begin{array}{lll}
\displaystyle{\hspace{-2mm}%
  \frac{C_{p,x}}{N}=\frac{25}{4}\Bigg\{ x\Bigg[\frac{\Phi_{5/2}^{(+)}(t_F)}{\Phi_{3/2}^{(+)}(t_F)} - \frac{3}{5}\frac{\Phi_{3/2}^{(+)}(t_F)}{\Phi_{1/2}^{(+)}(t_F)} \Bigg]\,+ %
}\vspace{2mm}\\ %
\displaystyle{%
\hspace{15mm} +\, (1-x)\Bigg[\frac{\Phi_{5/2}^{(-)}(t_B)}{\Phi_{3/2}^{(-)}(t_B)} - \frac{3}{5}\frac{\Phi_{3/2}^{(-)}(t_B)}{\Phi_{1/2}^{(-)}(t_B)} \Bigg]\!\Bigg\}\times   %
}\vspace{2mm}\\ %
\displaystyle{%
\hspace{7mm}\times
\frac{\displaystyle{\Bigg[\frac{6}{5}\frac{G}{nT}+x\frac{\Phi_{5/2}^{(+)}(t_F)}{\Phi_{3/2}^{(+)}(t_F)}+(1-x)\frac{\Phi_{5/2}^{(-)}(t_B)}{\Phi_{3/2}^{(-)}(t_B)}\Bigg]}}%
{\displaystyle{\Bigg[\frac{2G}{nT}+x\frac{\Phi_{3/2}^{(+)}(t_F)}{\Phi_{1/2}^{(+)}(t_F)}+(1-x)\frac{\Phi_{3/2}^{(-)}(t_B)}{\Phi_{1/2}^{(-)}(t_B)}\Bigg]}} %
}.%
\end{array}
\end{equation}
The parameter $\eta$ introduced in (\ref{53}) is determined by the
formula
\begin{equation} \label{56}
\begin{array}{ll}
\displaystyle{\hspace{0mm}%
  \eta=\frac{\displaystyle{\Bigg(3\beta G + 5 \frac{\Phi_{5/2}^{(+)}}{\Lambda_F^3} + \frac{5}{2}\frac{\Phi_{5/2}^{(-)}}{\Lambda_B^3}\Bigg)  }}
  {\displaystyle{\Bigg(2G + \frac{2T}{\Lambda_F^3}\frac{\Phi_{3/2}^{(+)\,2}}{\Phi_{1/2}^{(+)}}+\frac{T}{\Lambda_B^3}\frac{\Phi_{3/2}^{(-)\,2}}{\Phi_{1/2}^{(-)}}\Bigg)  }}.  %
}
\end{array}
\end{equation}
The designation is used
\begin{equation} \nonumber
\begin{array}{ll}
\displaystyle{\hspace{0mm}%
  G=g_F\big(n_F^2\big/4\big)+g_Bn_B^2+\,g_{F\!B}n_Fn_B=
}\vspace{2mm}\\ %
\displaystyle{\hspace{4mm}%
  = n^2\Big[g_F\big(x^2\big/4\big)+g_B(1-x)^2+g_{F\!B}\,x(1-x)\Big]. %
}
\end{array}
\end{equation}

The difference of the feat capacities (\ref{55}), (\ref{54})
\begin{equation} \label{57}
\begin{array}{lll}
\displaystyle{\hspace{-3mm}%
  \frac{C_{p,x}-C_{V,x}}{N}= %
}\vspace{2mm}\\ %
\displaystyle{%
\hspace{-2mm}  =\frac{25}{4}\Bigg\{
x\Bigg[\frac{\Phi_{5/2}^{(+)}(t_F)}{\Phi_{3/2}^{(+)}(t_F)} - \frac{3}{5}\frac{\Phi_{3/2}^{(+)}(t_F)}{\Phi_{1/2}^{(+)}(t_F)}\big]\,+ %
}\vspace{2mm}\\ %
\displaystyle{%
\hspace{12mm} +\, (1-x)\Bigg[\frac{\Phi_{5/2}^{(-)}(t_B)}{\Phi_{3/2}^{(-)}(t_B)} - \frac{3}{5}\frac{\Phi_{3/2}^{(-)}(t_B)}{\Phi_{1/2}^{(-)}(t_B)} \Bigg]\!\Bigg\}^2\times   %
}\vspace{2mm}\\ %
\displaystyle{%
\hspace{2mm}\times %
\frac{1}{\displaystyle{\Bigg[\frac{2G}{nT}+x\frac{\Phi_{3/2}^{(+)}(t_F)}{\Phi_{1/2}^{(+)}(t_F)}+(1-x)\frac{\Phi_{3/2}^{(-)}(t_B)}{\Phi_{1/2}^{(-)}(t_B)}\Bigg]}} %
}%
\end{array}
\end{equation}
should be positive in accordance with the requirement of
thermodynamic stability. The conditions of stability are analyzed
below. In an ideal Bose gas the heat capacity at a constant pressure
tends to infinity when approaching to the condensation temperature
from the side of higher temperatures, and in the phase with
condensate it proves to be infinite, since at $T<T_C$ in an ideal
gas the pressure does not depend on the density \cite{P2}. When the
interaction between Bose particles is taken into account the heat
capacity at a constant pressure at the transition temperature and
below it (see further section V) proves to be finite.

One more important characteristic of systems of interacting
particles is their compressibility. The isothermic compressibility
at a constant concentration $\gamma_{T,x}=n^{-1}(\partial n/\partial p)_{T,x}$, %
considering the conditions $dN_F=dN_B=dT=0$, is given by the formula
\begin{equation} \label{58}
\begin{array}{ll}
\displaystyle{\hspace{0mm}%
  \gamma_{T,x}=\frac{1}{\displaystyle{nT\Bigg[\frac{2G}{nT}+x\frac{\Phi_{3/2}^{(+)}(t_F)}{\Phi_{1/2}^{(+)}(t_F)}+(1-x)\frac{\Phi_{3/2}^{(-)}(t_B)}{\Phi_{1/2}^{(-)}(t_B)}\Bigg]}}. %
}
\end{array}
\end{equation}

The adiabatic compressibility $\gamma_{\sigma,x}=n^{-1}(\partial
n/\partial p)_{\sigma,x}$, where $\sigma=S\big/N$ is the entropy per
one particle of a solution, is obtained from the condition of
constancy of the entropy which is equivalent to the requirements
$t_F=const$, $t_B=const$ or $dt_F=dt_B=0$, that gives %
\begin{equation} \label{59}
\begin{array}{ll}
\displaystyle{\hspace{0mm}%
  \gamma_{\sigma,x}=\frac{1}{\displaystyle{nT\Bigg[\frac{2G}{nT}+\frac{5}{3}\,x\frac{\Phi_{5/2}^{(+)}(t_F)}{\Phi_{3/2}^{(+)}(t_F)}+\frac{5}{3}(1-x)\frac{\Phi_{5/2}^{(-)}(t_B)}{\Phi_{3/2}^{(-)}(t_B)}\Bigg]}}. %
}
\end{array}
\end{equation}
The relation proves to be satisfied
\begin{equation} \label{60}
\begin{array}{ll}
\displaystyle{\hspace{-2mm}%
\frac{\gamma_{T,x}}{\gamma_{\sigma,x}}=\frac{C_{p,x}}{C_{V,x}}= %
}\vspace{2mm}\\%
\displaystyle{\hspace{2mm}%
  =\frac{\displaystyle{\Bigg[\frac{2G}{nT}+\frac{5}{3}\,x\frac{\Phi_{5/2}^{(+)}(t_F)}{\Phi_{3/2}^{(+)}(t_F)}+\frac{5}{3}(1-x)\frac{\Phi_{5/2}^{(-)}(t_B)}{\Phi_{3/2}^{(-)}(t_B)}\Bigg]}}
   {\displaystyle{\Bigg[\frac{2G}{nT}+x\frac{\Phi_{3/2}^{(+)}(t_F)}{\Phi_{1/2}^{(+)}(t_F)}+(1-x)\frac{\Phi_{3/2}^{(-)}(t_B)}{\Phi_{1/2}^{(-)}(t_B)}\Bigg]}}. %
}
\end{array}
\end{equation}
Let us also give the formula for the square of speed of sound
$u_{\sigma,x}^2=1\big/[(m_Fn_F+m_Bn_B)\gamma_{\sigma,x}]$\,:  %
\begin{equation} \label{61}
\begin{array}{ll}
\displaystyle{\hspace{0mm}%
u_{\sigma,x}^2=T\frac{\displaystyle{\Bigg[\frac{2G}{nT}+\frac{5}{3}\,x\frac{\Phi_{5/2}^{(+)}(t_F)}{\Phi_{3/2}^{(+)}(t_F)}+\frac{5}{3}(1-x)\frac{\Phi_{5/2}^{(-)}(t_B)}{\Phi_{3/2}^{(-)}(t_B)}\Bigg]}}
   {\displaystyle{\big[x m_F + (1-x)m_B \big]}}. %
}
\end{array}
\end{equation}

\section{Thermodynamics of a mixture with Bose-Einstein condensate}
Bose-Einstein condensation in a system of interacting particles is
accompanied by the breaking of the phase symmetry of a state, which
manifests itself in the appearance of the anomalous quasiaverages
$\langle\Phi({\bf r})\rangle$, $\langle\Phi^+({\bf r})\rangle$ and %
$\langle\Phi({\bf r})\Phi({\bf r}')\rangle$, $\langle\Phi^+({\bf r})\Phi^+({\bf r}')\rangle$. %
The theoretical description of a system with taking into account the
breaking of the phase symmetry becomes considerably complicated
\cite{P3,P5}. The model of condensation of interacting particles
proposed in \cite{P1} accounts for Bose-Einstein condensation to the
extent that it takes place in an ideal gas, where the macroscopic
occupation of the ground state is taken into consideration but the
appearance of the quasiaverages $\langle\Phi({\bf r})\rangle$, $\langle\Phi^+({\bf r})\rangle$ %
associated with the one-particle condensate is disregarded. Although
in this approach one neglects the effects caused by the existence of
the anomalous quasiaverages, nevertheless, accounting for the
interaction enables to eliminate those difficulties which exist in
the model of an ideal gas \cite{P1}, and to make the model more
realistic. In this paper the phenomenon of condensation will be
considered in the same way as is done in \cite{P1} for a pure Bose
system.

Below the temperature of transition of the subsystem of Bose
particles into the phase with condensate $T<T_C$ the formulas of
thermodynamic of a mixture are obtained if one sets
$t_F=\beta\mu_{F*}$, $t_B=0$ in the above formulas relating to the
case $T>T_C$. Thus, the effective chemical potential of the Bose
subsystem proves to be fixed and equal to zero, and at that $t_{F0}<t_F<\infty$. %
The actual chemical potential of Bose particles remains a ``good''
thermodynamic variable, being different from zero, and is determined
by the formula
\begin{equation} \label{62}
\begin{array}{ll}
\displaystyle{\hspace{0mm}%
  \mu_B=2g_Bn_B+g_{F\!B}n_F.
}%
\end{array}
\end{equation}
The thermodynamic potential in the phase with condensate as a
function of the chemical potentials and temperature is given by the
formula
\begin{equation} \label{63}
\begin{array}{ll}
\displaystyle{\hspace{0mm}%
  \Omega=-V\Bigg[\frac{\big(g_Fg_B-g_{F\!B}^2\big)}{4g_B}\,n_F^2 + \frac{\mu_B^2}{4g_B}\, + %
}\vspace{1mm}\\ %
\displaystyle{%
  \hspace{28mm} + \frac{2T}{\Lambda_F^3}\Phi_{5/2}^{(+)}(t_F)+ \frac{T}{\Lambda_B^3}\Phi_{5/2}^{(-)}(0)\Bigg].   %
}
\end{array}
\end{equation}
It is easy to make sure that all necessary thermodynamic relations
are fulfilled in the phase with condensate:
$N_F=-(\partial\Omega/\partial\mu_F)_{V,T,\mu_B}$, $N_B=-(\partial\Omega/\partial\mu_B)_{V,T,\mu_F}$,  %
$S=-(\partial\Omega/\partial T)_{V,\mu_F,\mu_B}$. As seen, in the
phase with condensate in the thermodynamic potential the constant of
interaction of Bose particles stands in the denominator, so that
$\Omega$ is not an analytic function of the quantity $g_B$. This
indicates impossibility of using the perturbation theory in the
interaction constant $g_B$ in the phase with condensate. The
temperature dependence of the density of particles in the condensate
is the same as in a pure Bose system
\begin{equation} \label{64}
\begin{array}{ll}
\displaystyle{%
   n_0(T)=n_B\!\left[ 1- \left(\frac{T}{T_C}\right)^{\!3/2} \right], %
}
\end{array}
\end{equation}
where the condensation temperature is determined by the formula
(\ref{46}) and depends on the concentration of the admixture of
Fermi particles. The number density of overcondensate Bose particles
decreases with temperature
$\displaystyle{n_B'(T)=\frac{1}{\Lambda_B^3}\Phi_{3/2}^{(-)}(0)}$.

The pressure is given by the formula
\begin{equation} \label{65}
\begin{array}{ll}
\displaystyle{\hspace{0mm}%
  p=G + \frac{2T}{\Lambda_F^3}\Phi_{5/2}^{(+)}(t_F)+ \frac{T}{\Lambda_B^3}\Phi_{5/2}^{(-)}(0).%
} %
\end{array}
\end{equation}
The formula for the entropy remains the same as for the mixture of
ideal gases
\begin{equation} \label{66}
\begin{array}{ll}
\displaystyle{\hspace{0mm}%
  S=\frac{2V}{\Lambda_F^3}\Bigg[\frac{5}{2}\Phi_{5/2}^{(+)}(t_F)-t_F\Phi_{3/2}^{(+)}(t_F)\Bigg] + \frac{V}{\Lambda_B^3}\frac{5}{2}\Phi_{5/2}^{(-)}(0). %
}
\end{array}
\end{equation}

The heat capacities at constant volume and pressure, as well as at a
fixed concentration, have the form
\begin{equation} \label{67}
\begin{array}{ll}
\displaystyle{\hspace{-2mm}%
  \frac{C_{V,x}}{N}=\frac{15}{4}\Bigg\{ x\Bigg[\frac{\Phi_{5/2}^{(+)}(t_F)}{\Phi_{3/2}^{(+)}(t_F)} - \frac{3}{5}\frac{\Phi_{3/2}^{(+)}(t_F)}{\Phi_{1/2}^{(+)}(t_F)} \Bigg]\,+ %
}\vspace{2mm}\\ %
\displaystyle{%
\hspace{28mm} +\, (1-x)\left(\frac{T}{T_C}\right)^{\!\!3/2}\frac{\Phi_{5/2}^{(-)}(0)}{\Phi_{3/2}^{(-)}(0)} \!\Bigg\},   %
}
\end{array}
\end{equation}
\vspace{0mm}
\begin{equation} \label{68}
\begin{array}{ll}
\displaystyle{\hspace{-2mm}%
  \frac{C_{p,x}}{N}=\frac{25}{4}\Bigg\{ x\Bigg[\frac{\Phi_{5/2}^{(+)}(t_F)}{\Phi_{3/2}^{(+)}(t_F)} - \frac{3}{5}\frac{\Phi_{3/2}^{(+)}(t_F)}{\Phi_{1/2}^{(+)}(t_F)} \Bigg]\,+ %
}\vspace{2mm}\\ %
\displaystyle{%
\hspace{38mm} +\, (1-x)\left(\frac{T}{T_C}\right)^{\!\!3/2}\frac{\Phi_{5/2}^{(-)}(0)}{\Phi_{3/2}^{(-)}(0)} \!\Bigg\}\times   %
}\vspace{2mm}\\ %
\displaystyle{%
\hspace{5mm}\times
\frac{\displaystyle{\Bigg[\frac{6}{5}\frac{G}{nT}+x\frac{\Phi_{5/2}^{(+)}(t_F)}{\Phi_{3/2}^{(+)}(t_F)}+(1-x)\left(\frac{T}{T_C}\right)^{\!\!3/2}\frac{\Phi_{5/2}^{(-)}(0)}{\Phi_{3/2}^{(-)}(0)}  \Bigg]}}%
{\displaystyle{\Bigg[\frac{2G}{nT}+x\frac{\Phi_{3/2}^{(+)}(t_F)}{\Phi_{1/2}^{(+)}(t_F)}\Bigg]}} %
},%
\end{array}
\end{equation}
and their difference\,:
\begin{equation} \label{69}
\begin{array}{ll}
\displaystyle{\hspace{-3mm}%
\frac{C_{p,x}-C_{V,x}}{N}=
}\vspace{2mm}\\ %
\displaystyle{\hspace{-2mm}%
  =\frac{25}{4}\Bigg\{ x\Bigg[\frac{\Phi_{5/2}^{(+)}(t_F)}{\Phi_{3/2}^{(+)}(t_F)} - \frac{3}{5}\frac{\Phi_{3/2}^{(+)}(t_F)}{\Phi_{1/2}^{(+)}(t_F)} \Bigg]\,+ %
}\vspace{2mm}\\ %
\displaystyle{%
\hspace{-2mm} +\, (1-x)\!\left(\frac{T}{T_C}\right)^{\!\!3/2}\!\frac{\Phi_{5/2}^{(-)}(0)}{\Phi_{3/2}^{(-)}(0)} \!\Bigg\}^2\!  %
\frac{1}{\displaystyle{\Bigg[\frac{2G}{nT}+x\frac{\Phi_{3/2}^{(+)}(t_F)}{\Phi_{1/2}^{(+)}(t_F)}\Bigg]}}.
}
\end{array}
\end{equation}

The isothermic and adiabatic compressibilities are given by the
formulas:
\begin{equation} \label{70}
\begin{array}{ll}
\displaystyle{\hspace{0mm}%
  \gamma_{T,x}=\frac{1}{\displaystyle{nT\Bigg[\frac{2G}{nT}+x\frac{\Phi_{3/2}^{(+)}(t_F)}{\Phi_{1/2}^{(+)}(t_F)}\Bigg]}}, %
}
\end{array}
\end{equation}
\vspace{-4mm}
\begin{equation} \label{71}
\begin{array}{ll}
\displaystyle{\hspace{0mm}%
\gamma_{\sigma,x}=
}\vspace{2mm}\\ %
\displaystyle{\hspace{0mm}%
  =\!\frac{1}{\displaystyle{nT\Bigg[\frac{2G}{nT}+\!\frac{5}{3}\,x\frac{\Phi_{5/2}^{(+)}(t_F)}{\Phi_{3/2}^{(+)}(t_F)}+\!\frac{5}{3}(1-x)\!\left(\frac{T}{T_C}\right)^{\!\!3/2}\!\frac{\Phi_{5/2}^{(-)}(0)}{\Phi_{3/2}^{(-)}(0)}\Bigg]}}, %
}
\end{array}
\end{equation}
and the square of speed of first sound by the formula
\begin{equation} \label{72}
\begin{array}{ll}
\displaystyle{\hspace{0mm}%
u_{\sigma,x}^2=
}\vspace{2mm}\\ %
\displaystyle{\hspace{0mm}%
=T\frac{\displaystyle{\Bigg[\frac{2G}{nT}+\frac{5}{3}\,x\frac{\Phi_{5/2}^{(+)}(t_F)}{\Phi_{3/2}^{(+)}(t_F)}+\frac{5}{3}(1-x)\!\left(\frac{T}{T_C}\right)^{\!\!3/2}\!\frac{\Phi_{5/2}^{(-)}(0)}{\Phi_{3/2}^{(-)}(0)}\Bigg]}}
   {\displaystyle{\big[x m_F + (1-x)m_B \big]}}. %
}
\end{array}
\end{equation}

%\newpage
\section{Thermodynamic stability of a mixture of Fermi and Bose particles}
Now let us analyze the conditions of thermodynamic stability of the
spatially uniform state of solutions. As known \cite{LL}, in a
thermodynamically stable system the inequalities hold $C_p>C_V>0$, %
so that the difference of the heat capacities $C_p-C_V>0$ should be
positive. In the formulas (\ref{57}) and (\ref{69}) obtained above
the numerator on the right side is always positive, and the sign of
the denominator is significantly determined by the character of
interparticle interactions or, in the model being used, by the signs
of the scattering lengths of the mixture components and can have a
different sign. The requirements of thermodynamic stability are
satisfied, if at $T>T_C$ the function
\begin{equation} \label{73}
\begin{array}{ll}
\displaystyle{\hspace{0mm}%
  F_>\equiv\frac{2G}{nT}+x\frac{\Phi_{3/2}^{(+)}(t_F)}{\Phi_{1/2}^{(+)}(t_F)}+(1-x)\frac{\Phi_{3/2}^{(-)}(t_B)}{\Phi_{1/2}^{(-)}(t_B)}>0
}%
\end{array}
\end{equation}
is positive, and at $T<T_C$ the following one is positive
\begin{equation} \label{74}
\begin{array}{ll}
\displaystyle{\hspace{0mm}%
  F_<\equiv\frac{2G}{nT}+x\frac{\Phi_{3/2}^{(+)}(t_F)}{\Phi_{1/2}^{(+)}(t_F)}>0.
}%
\end{array}
\end{equation}
With the help of the formulas (\ref{38}), (\ref{50}), (\ref{65}) it
is easy to make sure that the inequalities (\ref{73}) and (\ref{74})
are equivalent to the known thermodynamic inequality $(\partial p/\partial V)_{T,x}<0$ \cite{LL}. %
The relations (\ref{73}) and (\ref{74}) can be represented in the form %
\begin{equation} \label{73v}
\begin{array}{ll}
\displaystyle{\hspace{0mm}%
  \frac{1}{2}\Bigg[g_F+\frac{T\Lambda_F^3}{\Phi_{1/2}^{(+)}(t_F)}\Bigg]n_F^2\, + %
  \Bigg[2g_B+\frac{T\Lambda_B^3}{\Phi_{1/2}^{(-)}(t_B)}\Bigg]n_B^2 + %
}\vspace{2mm}\\%
\displaystyle{\hspace{47mm}%
  +\, 2g_{F\!B}n_Fn_B>0,
}
\end{array}
\end{equation}
\vspace{-3mm}
\begin{equation} \label{74v}%
\begin{array}{ll}
\displaystyle{\hspace{-2mm}%
  \frac{1}{2}\Bigg[g_F+\!\frac{T\Lambda_F^3}{\Phi_{1/2}^{(+)}(t_F)}\Bigg]n_F^2 + %
  2g_Bn_B^2 + 2g_{F\!B}n_Fn_B>0. %
}
\end{array}
\end{equation}
The conditions of positive definiteness of the quadratic forms (\ref{73v}), (\ref{74v}) %
are: at $T>T_C$
\begin{equation} \label{75v}
\begin{array}{ll}
\displaystyle{\hspace{0mm}%
  \Bigg[g_B+\frac{T\Lambda_B^3}{2\Phi_{1/2}^{(-)}(t_B)}\Bigg]\!\Bigg[g_F+\frac{T\Lambda_F^3}{\Phi_{1/2}^{(+)}(t_F)}\Bigg]-g_{F\!B}^2>0, %
}\vspace{2mm}\\%
\displaystyle{\hspace{0mm}%
  g_B+\frac{T\Lambda_B^3}{2\Phi_{1/2}^{(-)}(t_B)}>0\quad{\textit{or}}\quad  g_F+\frac{T\Lambda_F^3}{\Phi_{1/2}^{(+)}(t_F)}>0, %
}
\end{array}
\end{equation}
and at $T<T_C$
\begin{equation} \label{76v}
\begin{array}{ll}
\displaystyle{\hspace{0mm}%
  g_B\Bigg[g_F+\frac{T\Lambda_F^3}{\Phi_{1/2}^{(+)}(t_F)}\Bigg]-g_{F\!B}^2>0, %
}\vspace{2mm}\\%
\displaystyle{\hspace{0mm}%
  g_B>0\quad{\textit{or}}\quad  g_F+\frac{T\Lambda_F^3}{\Phi_{1/2}^{(+)}(t_F)}>0. %
}
\end{array}
\end{equation}

In the analysis of stability and in specific calculations of
thermodynamic quantities it is convenient to pass to the
dimensionless form of their representation. Let us introduce some
characteristic length $l_0$, which can be defined by reasons of
convenience depending on the value of the density of a considered
system. Also, let us define the characteristic temperature
\begin{equation} \label{75}
\begin{array}{ll}
\displaystyle{\hspace{0mm}%
  T_*\equiv\frac{2\pi\hbar^2}{m_Bl_0^2}\frac{1}{\big[\Phi_{3/2}^{(-)}(0)\big]^{2/3}}. %
}%
\end{array}
\end{equation}
The dimensionless density and temperature are defined by the formulas: %
$\tilde{n}=nl_0^3$, $\tilde{T}=T\big/T_*$. The introduction of the
characteristic temperature by the formula (\ref{75}) allows to write
the dimensionless temperature of Bose-Einstein condensation in the form %
$\tilde{T}_C=\tilde{T}_{C0}(1-x)^{2/3}$, where
$\tilde{T}_{C0}=\tilde{n}^{2/3}$ is the dimensionless condensation
temperature of a pure Bose system. The interaction constants can be
expressed through the scattering lengths $a_F,\,a_B,\,a_{F\!B}$, %
which in the Born approximation are determined by the formulas:
\begin{equation} \label{76}
\begin{array}{ll}
\displaystyle{\hspace{0mm}%
  g_F=\frac{4\pi\hbar^2}{m_F}a_F, \quad g_B=\frac{4\pi\hbar^2}{m_B}a_B, \quad g_{F\!B}=\frac{2\pi\hbar^2}{m_{F\!B}}a_{F\!B},  %
}%
\end{array}
\end{equation}
where $m_{F\!B}=m_Bm_F\big/(m_B+m_F)$ is the reduced mass.
Everywhere it is assumed that the repulsion $a_B>0$ dominates
between Bose particles, that ensures the stability of the pure
subsystem of Bose particles. It is also convenient to introduce the
ratios of the
scattering lengths $\alpha_F\equiv a_F/a_B$, $\alpha_{F\!B}\equiv a_{F\!B}/a_B$ %
and the ratio of the masses $\tilde{m}\equiv m_F/m_B$,
$m_{F\!B}/m_B=\tilde{m}/(1+\tilde{m})$. In these designations %
$G\big/nT\equiv\big(\tilde{n}\big/\tilde{T}\big)Z(x)$, where
\begin{equation} \hspace{0mm}\label{77}
\begin{array}{ll}
\displaystyle{\hspace{0mm}%
  Z(x)\equiv 2\tilde{a}_B\big[\Phi_{3/2}^{(-)}(0)\big]^{2/3}\times %
}\vspace{1mm}\\ %
\displaystyle{\hspace{0mm}%
\times\Bigg[(1-x)^2 + \frac{\alpha_F}{\tilde{m}}\frac{x^2}{4}+\frac{\alpha_{F\!B}}{2}\big(1+\tilde{m}^{-1}\big)\,x(1-x)\Bigg]   %
}
\end{array}
\end{equation}
and $\tilde{a}_B=a_B/l_0$. When using the introduced designations
the condition of stability of the phase without condensate
(\ref{73}) takes the form
\begin{equation} \label{78}
\begin{array}{ll}
\displaystyle{\hspace{0mm}%
  F_>(\tilde{n},\tilde{T},x;t_F,t_B)\equiv
}\vspace{2mm}\\ %
\displaystyle{\hspace{1mm}%
  \equiv \frac{2\tilde{n}}{\tilde{T}}Z(x)+x\frac{\Phi_{3/2}^{(+)}(t_F)}{\Phi_{1/2}^{(+)}(t_F)}+(1-x)\frac{\Phi_{3/2}^{(-)}(t_B)}{\Phi_{1/2}^{(-)}(t_B)}>0,
}%
\end{array}
\end{equation}
and the relations following from (\ref{37}) should be taken into account %
\begin{equation} \label{79}
\begin{array}{ll}
\displaystyle{\hspace{0mm}%
  x=2\tilde{m}^{3/2}\frac{\tilde{T}^{3/2}}{\tilde{n}}\frac{\Phi_{3/2}^{(+)}(t_F)}{\Phi_{3/2}^{(-)}(0)},\,\,\,\,\,
  1-x=\frac{\tilde{T}^{3/2}}{\tilde{n}}\frac{\Phi_{3/2}^{(-)}(t_B)}{\Phi_{3/2}^{(-)}(0)}.
}
\end{array}
\end{equation}
In the phase with condensate
\begin{equation} \label{80}
\begin{array}{ll}
\displaystyle{\hspace{0mm}%
  F_<(\tilde{n},\tilde{T},x;t_F)\equiv \frac{2\tilde{n}}{\tilde{T}}Z(x)+x\frac{\Phi_{3/2}^{(+)}(t_F)}{\Phi_{1/2}^{(+)}(t_F)}>0,
}
\end{array}
\end{equation}
where the first formula in (\ref{79}) should be taken into account.
On the right side of the relations (\ref{78}),(\ref{80}) only the
function $Z(x)$ can be negative. According to data given in
\cite{Jamieson}, the scattering lengths for $^3$He\,--\,$^3$He and
$^3$He\,--\,$^4$He are negative, and the scattering length for
$^4$He\,--\,$^4$He is positive. In specific calculations we assume
$a_B=20\,\textrm{{\AA}}$, $a_F=-10\,\textrm{{\AA}}$, $a_{F\!B}=-10\,\textrm{{\AA}}$. %
The characteristic length is assumed to be
$l_0=n_0^{-1/3}=10^{-6}\,\textrm{cm}=100\,\textrm{{\AA}}$, which
corresponds to the density $n_0=10^{18}\,\textrm{cm}^{-3}$. The
ratio of the masses $\tilde{m}\equiv m_F/m_B=3/4$. The concentration
$x_c$ is determined by the condition $Z(x_c)=0$ and at chosen
parameters $x_c=0.56$. The concentration $x_0$ separates the regions
of the uniform and nonuniform states at zero temperature and is
determined by the equation
$\displaystyle{Z(x_0)+\frac{x_0^{5/3}}{4\tilde{m}}\Bigg[\!\frac{\pi\zeta^2(3/2)}{3\tilde{n}}\!\Bigg]^{\!1/3} }=0$. %
This same equation determines the density $\tilde{n}_0$ at a fixed
concentration (Fig.\,2). In Figs.\,1\,--\,3 there are shown the
region of the spatially uniform state and the region of
nonuniformity in different coordinates.

\vspace{0mm} %
\begin{figure}[t!]
\vspace{-0mm}  %\hspace{0mm}
\includegraphics[width = 1.01\columnwidth]{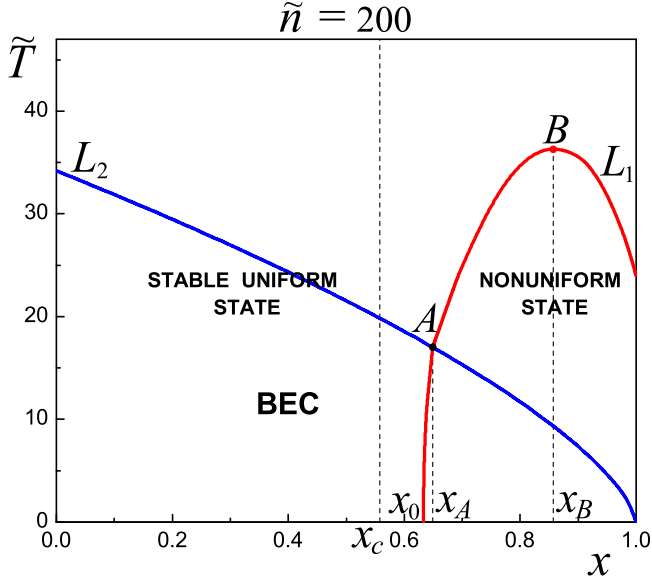} % 1.0\columnwidth - КОНТРОЛЬ
\vspace{-8mm} % \vspace{-8mm}
\caption{\label{fig01} %
The diagram of stability of a mixture in the coordinates temperature
-- concentration at $\tilde{n}=200$: \newline %
$L_1$ -- the boundary of the region of the stable uniform state, \newline %
$L_2$ -- the condensation line $\tilde{T}_C(x)\equiv\tilde{n}^{2/3}(1-x)^{2/3}$, %
$x_0=0.63$, $A=(x_A=0.65,\tilde{T}_A=17.0)$ -- the intersection point of $L_1,L_2$, %
$B=(x_B=0.86,\tilde{T}_B=36.3)$ -- the maximum point of $L_1$. %
}%
\end{figure}
\vspace{0mm} %
\begin{figure}[t!]
\vspace{0.5mm}  %\hspace{0mm}
\includegraphics[width = 0.99\columnwidth]{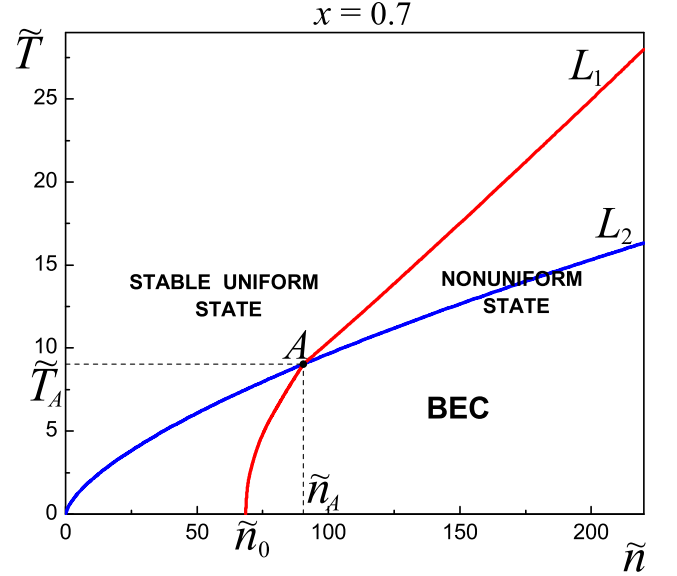} % 1.0\columnwidth - КОНТРОЛЬ
\vspace{-8mm} % \vspace{-8mm}
\caption{\label{fig02} %
The diagram of stability of a mixture in the coordinates temperature
-- density at $x=0.7$: \newline %
$L_1$ -- the boundary of the region of the stable uniform state, \newline %
$L_2$ -- the condensation line $\tilde{T}_C(\tilde{n})\equiv\tilde{n}^{2/3}(1-x)^{2/3}$, %
$\tilde{n}_0=68.5$, $A=(\tilde{n}_A=90.5,\tilde{T}_A=9.03)$ -- the intersection point of $L_1,L_2$. %
\newline \vspace{2mm} %
}%
\end{figure}
\vspace{0mm} %
\begin{figure}[b!]
\vspace{-6mm}  %\hspace{0mm}
\includegraphics[width = 0.99\columnwidth]{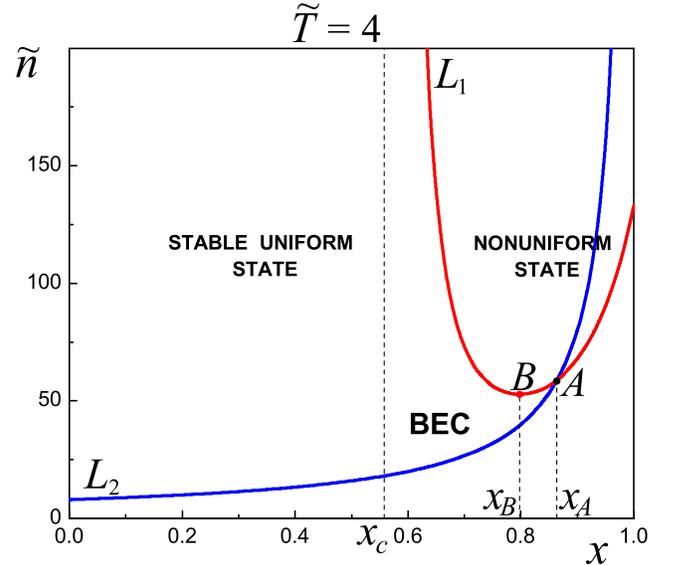} % 1.0\columnwidth - КОНТРОЛЬ
\vspace{-8mm} % \vspace{-8mm}
\caption{\label{fig03} %
The diagram of stability of a mixture in the coordinates density -- concentration at $\tilde{T}=4.0$: \newline %
$L_1$ -- the boundary of the region of the stable uniform state, \newline %
$L_2$ -- the condensation line $\tilde{n}_C(x)=\tilde{T}^{3/2}\!\big/(1-x)$, \newline%
$A=(x_A=0.86,\tilde{n}_A=58.5)$ -- the intersection point of $L_1,L_2$,\newline %
$B=(x_B=0.80,\tilde{n}_B=52.7)$ -- the minimum point of $L_1$.
}%
\end{figure}

In Fig.\,1 is shown the diagram of stability of a mixture in the
coordinates $(x,\tilde{T})$. At the concentrations lower than $x_0$
the spatially uniform state proves to be stable at all temperatures.
At $x_0<x<x_A$ the uniform state without condensate is stable at all
temperatures, and the state with condensate is stable only just
below the condensation line and becomes unstable with decreasing
temperature. At $x>x_A$ there is the region of stability of the
state without condensate (above the line $L_1$ in Fig.\,1), and the
state with condensate is always unstable. With decreasing the
density the region of instability diminishes and the point $x_0$
shifts to the right. Thus, at $\tilde{n}=100$ the concentration $x_0=0.67$. %

In Fig.\,2 is shown the diagram of stability in the  % of a mixture
coordinates $(\tilde{n},\tilde{T})$. At the low densities $\tilde{n}<\tilde{n}_0$ %
the spatially uniform state is stable at all temperatures. At
$\tilde{n}_0<\tilde{n}<\tilde{n}_A$ the uniform state without
condensate is also stable, and the state with condensate is stable
just below the condensation line and loses stability with decreasing
temperature. At $\tilde{n}>\tilde{n}_A$ the state with condensate is
always unstable, and the state without condensate is stable at high
temperatures (above the line $L_1$ in Fig.\,2).
%\newpage\vspace{0mm}

In Fig.\,3 is shown the diagram of stability in the coordinates
$(x,\tilde{n})$. At low concentrations $x<x_c$ the spatially uniform
state is stable at all densities. At $x_c<x<x_A$ the state
without condensate is stable, and the spatially uniform state % becomes unstable
loses stability with increasing the density. At $x>x_A$ the state
with condensate is unstable and there is only the region of
stability of the state without condensate at low densities (below
the line $L_1$ in Fig.\,3).

Thus, the stability of the spatially uniform state of a mixture
improves with decreasing the density and the concentration and with
increasing the temperature. At densities and concentrations lower
than some critical values, which are determined by the magnitude of
interactions, the system becomes stable at all temperatures.

In the particular case of zero temperature the stability of a
Fermi-Bose mixture was analyzed in work \cite{Viverit}. %

\begin{figure}[b!]
\vspace{-1mm}  %\hspace{0mm}
\includegraphics[width = 0.99\columnwidth]{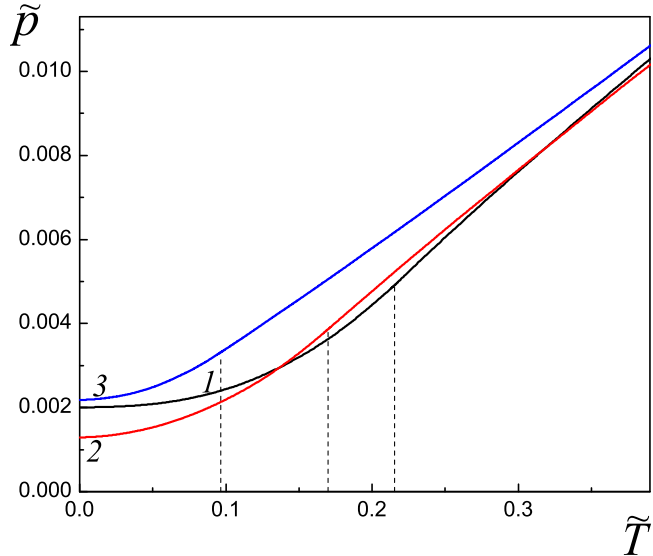} % 1.0\columnwidth - КОНТРОЛЬ
\vspace{-8mm} % \vspace{-8mm}
\caption{\label{fig04} %
The dependencies of the total pressure $\tilde{p}=p/p_*$ $(p_*\equiv 4\pi\hbar^2\!\big/m_B l_0^5)$ %
on the temperature $\tilde{T}$ at the density $\tilde{n}=0.1$ $(n=10^{17}\,\textrm{cm}^{-3})$ %
and various values of concentration $x$: %
({\it 1}) the pure Bose system $x=0$, $\tilde{T}_C=0.22$; %
({\it 2}) $x=0.3$, $\tilde{T}_C=0.17$; %
({\it 3}) $x=0.7$, $\tilde{T}_C=0.10$. %
The vertical dashed lines denote the condensation temperature.
}%
\end{figure}
\begin{figure}[t!]
\vspace{-0mm}  %\hspace{0mm}
\includegraphics[width = 0.99\columnwidth]{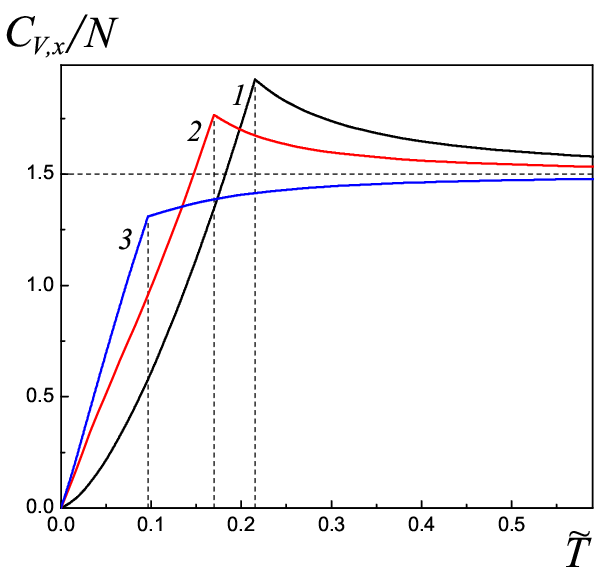} % 1.0\columnwidth - КОНТРОЛЬ
\vspace{-8mm} % \vspace{-8mm}
\caption{\label{fig05} %
The temperature dependencies of the heat capacity at a constant
volume at $\tilde{n}=0.1$ $(n=10^{17}\,\textrm{cm}^{-3})$ and various concentrations: %
({\it 1}) the pure Bose system $x=0$, $\tilde{T}_C=0.22$; %
({\it 2}) $x=0.3$, $\tilde{T}_C=0.17$; %
({\it 3}) $x=0.7$, $\tilde{T}_C=0.10$. %
The vertical dashed lines denote the condensation temperature.
}%
\end{figure}
\begin{figure}[h!]
\vspace{-0mm}  %\hspace{0mm}
\includegraphics[width = 0.99\columnwidth]{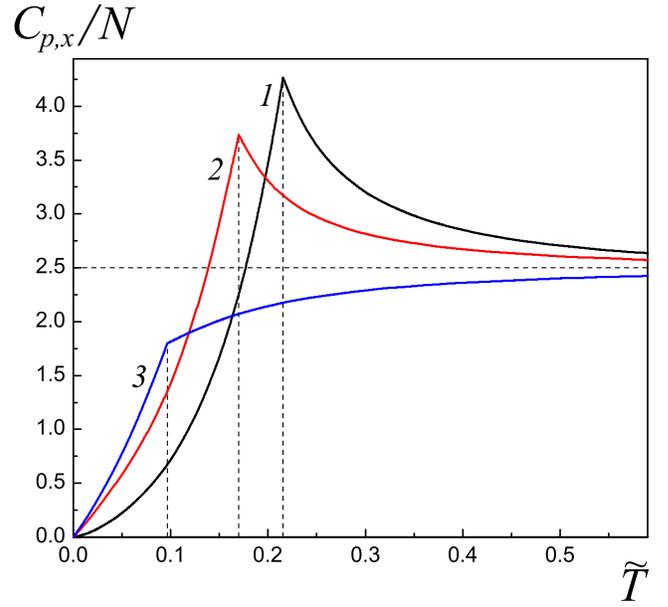} % 1.0\columnwidth
\vspace{-8mm} % \vspace{-8mm}
\caption{\label{fig06} %
The temperature dependencies of the heat capacity at a constant
pressure at $\tilde{n}=0.1$ $(n=10^{17}\,\textrm{cm}^{-3})$ and various concentrations: %
({\it 1}) the pure Bose system $x=0$, $\tilde{T}_C=0.22$; %
({\it 2}) $x=0.3$, $\tilde{T}_C=0.17$; %
({\it 3}) $x=0.7$, $\tilde{T}_C=0.10$. %
The vertical dashed lines denote the condensation temperature.
}%
\end{figure}

\section{Temperature dependencies of pressure, heat capacities and
compressibilities} %
In this section we present the results of some numerical
calculations based on the general formulas obtained in the previous
sections for the equation of state, temperature dependencies of heat
capacities, compressibilities and also the speed of sound. %
Being of the most interest, the behavior of thermodynamic quantities
near the temperature of transition into the state with condensate
can be analyzed if we take advantage of the expansions \vspace{-5mm} %
\begin{equation} \label{81}
\begin{array}{cc}
\displaystyle{\hspace{0mm}%
\Phi_{1/2}^{(-)}(t)\approx\!\sqrt{-\pi/t}+\zeta(1/2),\,\,\, \Phi_{3/2}^{(-)}(t)\approx\zeta(3/2)-2\sqrt{-\pi t}, %
}\vspace{4mm}\\ %
\displaystyle{\hspace{0mm}%
\Phi_{5/2}^{(-)}(t)\approx\zeta(5/2)+\zeta(3/2)\,t, %
}
\end{array}
\end{equation}
being valid for $|t|\ll 1$.

In Fig.\,4 are presented the graphs of dependencies of the total
pressure on temperature at a constant density and some concentrations. %
The pressure monotonically increases with increasing temperature. %
At the point of Bose-Einstein transition both the pressure and its
first derivative remain continuous. Only the second derivative
undergoes a jump
$\Delta\big(\partial^2 p\big/\partial T^2\big)\equiv\big(\partial^2 p\big/\partial T^2\big)_{T_C+0}-\big(\partial^2 p\big/\partial T^2\big)_{T_C-0}$\,: %
\begin{equation} \label{82}
\begin{array}{ll}
\displaystyle{\hspace{0mm}%
\Delta\!\left(\!\frac{\partial^2 p}{\partial T^2}\!\right)=-\frac{9}{8\pi}\frac{\big[\zeta(3/2)\big]^3}{T_C\big[\Lambda_B(T_C)\big]^3} = %
}\vspace{2mm}\\ %
\displaystyle{\hspace{15mm}%
=-\frac{9}{8\pi}\big[\zeta(3/2)\big]^3\left(\frac{m_B}{2\pi\hbar^2}\right)^{\!3/2}T_C^{1/2}. %
}
\end{array}
\end{equation}
Since the condensation temperature decreases with increasing the
concentration, then the value of the jump $\Delta\big(\partial^2 p\big/\partial T^2)$ %
decreases respectively. Note that the pressure of a mixture at
$x=0.3$ (curve 2 in Fig.\,4) at zero temperature proves to be lower
than the pressure of a pure Bose system, which is conditioned by the
attraction between fermions and between fermions and bosons.

The numerical calculation of the temperature dependencies of the
heat capacities is presented in Figs.\,5 and 6. In Fig.\,5 are shown
the temperature dependencies of the heat capacity per one particle
at a constant volume for some values of the concentration. %
In a pure Bose system and at low concentrations the capacity at the
condensation temperature has a sharp maximum (curves 1,\,2), and at
greater concentrations (curve 3) the maximum is absent but there
remains the break. The form of the temperature dependencies of the
isobaric heat capacity $C_{p,x}(\tilde{T})\big/N$, shown in
Fig.\,6, is qualitatively similar to the dependencies $C_{V,x}(\tilde{T})\big/N$, %
but the peaks at the condensation temperature at low concentrations
prove to be more sharp.

\begin{figure}[b!]
\vspace{-1mm}  %\hspace{0mm}
\includegraphics[width = 0.99\columnwidth]{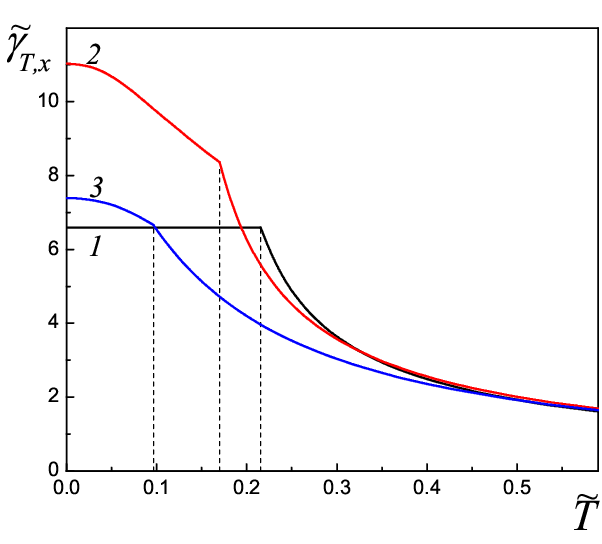} % 1.0\columnwidth - КОНТРОЛЬ
\vspace{-8mm} % \vspace{-8mm}
\caption{\label{fig07} %
The temperature dependencies of the isothermal compressibility
$\tilde{\gamma}_{T,x}=\gamma_{T,x}nT_*$ at %
$\tilde{n}=0.1$ $(n=10^{17}\,\textrm{cm}^{-3})$ and various concentrations: %
({\it 1}) the pure Bose system $x=0$, $\tilde{T}_C=0.22$; %
({\it 2}) $x=0.3$, $\tilde{T}_C=0.17$; %
({\it 3}) $x=0.7$, $\tilde{T}_C=0.10$. %
The vertical dashed lines denote the condensation temperature.
}%
\end{figure}
\begin{figure*}[t!]
\vspace{-2mm}  %\hspace{0mm}
\includegraphics[width = 1.0\textwidth]{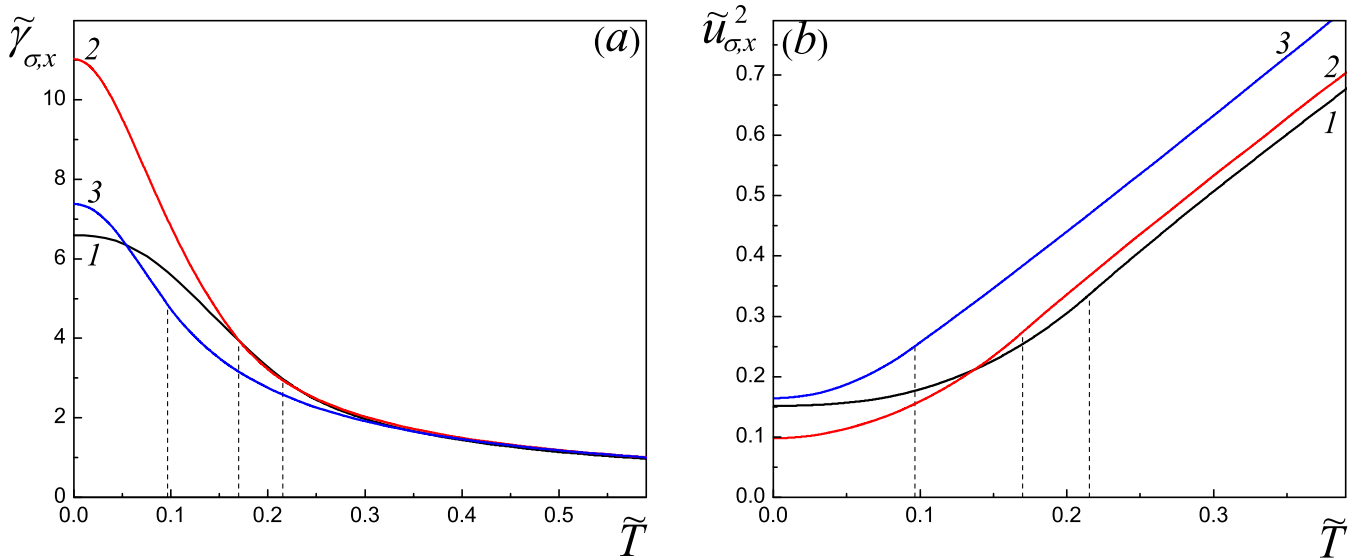} % 1.0\columnwidth
\vspace{-8mm} % \vspace{-8mm}
\caption{\label{fig08} %
The temperature dependencies of ({\it a}) the adiabatic compressibility %
$\tilde{\gamma}_{\sigma,x}=\gamma_{\sigma,x}nT_*$ %
and ({\it b}) the square of speed of sound $\tilde{u}_{\sigma,x}^2=u_{\sigma,x}^2(m_B/T_*)$ %
at $\tilde{n}=0.1$ $(n=10^{17}\,\textrm{cm}^{-3})$ and various concentrations: %
({\it 1}) the pure Bose system $x=0$, $\tilde{T}_C=0.22$; %
({\it 2}) $x=0.3$, $\tilde{T}_C=0.17$; %
({\it 3}) $x=0.7$, $\tilde{T}_C=0.10$. %
The vertical dashed lines denote the condensation temperature.
}%
\end{figure*}

The decrease of the magnitudes of the heat capacity peaks at the
condensation temperature as the concentration of fermions increases
is qualitatively similar to what is observed in $^3$He\,--\,$^4$He
solutions (see \cite{Eselson}, ch. 2). Both heat capacities are
continuous at the condensation temperature, but their derivatives at
the transition from the high-temperature to the low-temperature
phase undergo a jump
$\Delta\big(\partial C\big/\partial T\big)\equiv\big(\partial C\big/\partial T\big)_{T_C+0}-\big(\partial C\big/\partial T\big)_{T_C-0}$\,: %
\begin{equation} \label{83}
\vspace{04mm}
\begin{array}{ll}
\displaystyle{\hspace{0mm}%
\Delta\!\left(\!\frac{\partial C_{V,x}}{\partial T}\!\right)=-(1-x)\frac{27}{16\pi}\zeta^2(3/2)\frac{N}{T_C}\,, %
}\vspace{5mm}\\ %
\displaystyle{\hspace{00mm}%
\Delta\!\left(\!\frac{\partial C_{p,x}}{\partial T}\!\right)=-(1-x)\frac{27}{16\pi}\zeta^2(3/2)\frac{N}{T_C}\times %
}\vspace{2mm}\\ %
\displaystyle{\hspace{04mm}%
\times\!
\left[\frac{\displaystyle{1+\frac{5}{6}\frac{nT_C}{G}x\frac{\Phi_{5/2}^{(+)}(t_{F0})}{\Phi_{3/2}^{(+)}(t_{F0})}+\!\frac{5}{6}\frac{nT_C}{G}(1-x)\frac{\zeta(5/2)}{\zeta(3/2)}}}
{\displaystyle{1+\frac{nT_C}{2G}x\frac{\Phi_{3/2}^{(+)}(t_{F0})}{\Phi_{1/2}^{(+)}(t_{F0})}  }}\right]^2\!. %
}
\end{array}
\vspace{00mm}
\end{equation}
At $x=0$ these formulas, naturally, turn into the formulas of work \cite{P1}. %
In the limit of strong interaction $G\big/nT_C\gg 1$ the jumps of
derivatives for both heat capacities (\ref{83}) coincide. In the
opposite limit of weak interaction $G\big/nT_C\ll 1$ the jump of the
isobaric heat capacity proves to be independent of the interaction
constants: \vspace{01mm}
\begin{equation} \label{84}
\begin{array}{ll}
\displaystyle{\hspace{00mm}%
\Delta\!\left(\!\frac{\partial C_{p,x}}{\partial T}\!\right)=-\frac{(1-x)}{x^2}\frac{27}{16\pi}\zeta^2(3/2)\frac{N}{T_C}\times %
}\vspace{2mm}\\ %
\displaystyle{\hspace{04mm}%
\times\!
\left[\frac{5}{3}\frac{\Phi_{1/2}^{(+)}(t_{F0})}{\Phi_{3/2}^{(+)}(t_{F0})}\Bigg(
\displaystyle{x\frac{\Phi_{5/2}^{(+)}(t_{F0})}{\Phi_{3/2}^{(+)}(t_{F0})}+(1-x)\frac{\zeta(5/2)}{\zeta(3/2)}}  %
\Bigg)\right]^2\!. %
}
\end{array}
\vspace{06mm}
\end{equation}
In this respect a mixture of Fermi and Bose particles differs from a
pure Bose system, where at $G\rightarrow 0$ the heat capacity $C_p$
and the jump $\Delta\big(\partial C_p\big/\partial T\big)$ tend to
infinity.

In Figs.\,7 and 8 are shown the temperature dependencies of the
isothermal and adiabatic compressibilities and the square of speed of sound. %
In a pure Bose system of interacting particles below the
condensation temperature the isothermal compressibility proves to be
independent of temperature (curve 1 in Fig.\,7) \cite{P1}. %
The presence of the admixture of Fermi particles leads to the
appearance of dependence of the isothermal compressibility on
temperature in the state with condensate as well (curve 2,\,3 in Fig.\,7). %
At the condensation temperature the derivative with respect to
temperature of the isothermal compressibility undergoes a jump:
\begin{equation} \label{85}
\begin{array}{ll}
\displaystyle{\hspace{0mm}%
\Delta\!\left(\!\frac{\partial \gamma_{T,x}}{\partial T}\!\right)\!\!=\!-\frac{3n(1\!-x)\zeta^2\!(3/2)}{16\pi G^2} %
\Bigg[1\!+x\frac{nT_C}{2G}\frac{\Phi_{3/2}^{(+)}(t_{F0})}{\Phi_{1/2}^{(+)}(t_{F0})}\Bigg]^{\!-2}\!. %
}
\end{array}
\vspace{02mm}
\end{equation}
The adiabatic compressibility and the speed of sound, as well as
their first derivatives are continuous at the condensation
temperatures (Fig.\,$8a,b$).

\vspace{0mm}
\newpage
\section{Conclusion}\vspace{-1mm}
In the paper, in general form for the pair potentials of the
interparticle interaction,  there are formulated the self-consistent
field equations and obtained the thermodynamic relations for a
mixture of Bose and Fermi particles. The case of a delta-like
interaction between particles is studied in detail. %
Formulas for the thermodynamic potential, entropy, pressure, heat
capacities at constant volume and pressure, isothermal and adiabatic
compressibilities, speed of sound are obtained both above the
temperature of Bose-Einstein condensation and in the state with
condensate. The results of numerical calculations of these
quantities as functions of temperature at different concentrations
are presented. It is shown that with increasing the concentration of
the admixture of Fermi particles, the temperature of Bose-Einstein
condensation decreases and the features of thermodynamic quantities
at the transition temperature, in particular of the heat capacity,
become less pronounced. As in the case of a pure Bose system
\cite{P1}, in the phase with condensate the dependence of
thermodynamic quantities on the interaction constant between Bose
particles proves to be nonanalytic, so that developing the
perturbation theory in the interaction constant proves to be
impossible here.

%\vspace{50mm}

\end{document}